\documentclass[sigconf]{acmart}

\setcopyright{none}
\usepackage{url}

\usepackage{dcolumn}
\newcolumntype{d}{D{.}{.}{-1}} 
\newcolumntype{e}{D{E}{E}{-1}} 

\usepackage{tikz}
\usepackage{amsmath}

\usepackage{etoolbox}

\usepackage[utf8]{inputenc}
\usepackage[T1]{fontenc} 

\usepackage[english]{babel} 

\usepackage[inline]{enumitem}
\usepackage{xspace}

\newcommand{\mypara}[1]{\vspace{5pt}\noindent{\textbf{#1}}}

\usepackage{graphicx}
\usepackage{blindtext} 
\usepackage{longtable}

\usepackage{makecell} 
\usepackage{enumitem} 

\usepackage{dashrule}
\usepackage{subcaption}
\usepackage{placeins}
\usepackage[ruled,vlined]{algorithm2e}

\usepackage{listings}
\usepackage[most]{tcolorbox}
\definecolor[named]{gray}{rgb}{0.5,0.5,0.5}
\definecolor[named]{darkBlue}{rgb}{0,0,0.5}
\definecolor[named]{darkViolet}{rgb}{0.5,0,0.5}
\definecolor[named]{blueGreen}{rgb}{0,0.5,0.5}
\definecolor[named]{commentgreen}{rgb}{0,0.5,0}
\definecolor[named]{orange}{rgb}{0.61,0,0}
\definecolor[named]{blue}{rgb}{0.275,0.396,0.635}
\definecolor[named]{lightBlue}{rgb}{0,0.4,0.8}
\definecolor[named]{greenYellow}{rgb}{0.4,0.1,0}

\lstdefinelanguage{mycpp}{
    language=c++,
    deletekeywords={or},
    morekeywords=[1]{override, uint64_t},
    keywordstyle={[1]\color{darkBlue}\ttfamily},
    morekeywords=[2]{continue, break, in, config},
    keywordstyle={[2]\color{darkViolet}\ttfamily},
    commentstyle=\color{commentgreen}\ttfamily,
    stringstyle=\color{orange},
    numberstyle=\scriptsize\color{gray},
    rulecolor=\color{gray},
}
\lstnewenvironment{cpp}[1][]{\lstset{
    language=mycpp,
    basicstyle=\small\ttfamily,
    #1
}}{}
\lstdefinelanguage{wasm}{
  sensitive=true,
  otherkeywords={.add},
  morekeywords=[1]{add, sub, mul, div_s, const, gt_s, lt_s, ge_s, le_s, eq, ne, eqz, mut, tee, get, set, call, call_indirect, load, store}
  keywordstyle={[1]\color{darkBlue}\bfseries\ttfamily},
  morekeywords=[2]{i32,f32,i64,f64},
  keywordstyle={[2]\color{darkViolet}\bfseries\ttfamily},
  morekeywords=[3]{param, result, global,local, memory },
  keywordstyle={[3]\color{blueGreen}\bfseries\ttfamily},
  morekeywords=[5]{module, func, import, export, data, elem, table, type, elem, end, if, else},
  keywordstyle={[5]\color{darkBlue}\bfseries\ttfamily},
  morekeywords=[7]{(,),[,],.},
  keywordstyle={[7]\color{black}},
  numberstyle=\scriptsize\color{gray},
  rulecolor=\color{gray},
  morecomment=[l][\itshape\color{commentgreen}]{;;},
}

\lstnewenvironment{webassembly}[1][]{\lstset{
    language=wasm,
    basicstyle=\small\ttfamily,
    keywordstyle=\color{darkBlue}\bfseries,
    #1
}}{}

\lstdefinelanguage{wql}{
  language=c,
  sensitive=true,
  morekeywords=[1]{return, function, foreach, if, else, while, nil, true, false},
  keywordstyle={[1]\color{darkBlue}\ttfamily},
  morekeywords=[2]{continue, break, in, config},
  keywordstyle={[2]\color{darkViolet}\ttfamily},
  morekeywords=[3]{label, instType, name},
  keywordstyle={[3]\color{blueGreen}\ttfamily},
  stringstyle=\ttfamily\color{orange},
  showstringspaces=false,
  comment=[l]{//},
  morecomment=[s]{/*}{*/},
  commentstyle=\color{commentgreen}\ttfamily,
  numberstyle=\scriptsize\color{gray},
  rulecolor=\color{gray},
  tabsize=2,
}

\lstnewenvironment{wql}[1][]{\lstset{
    language=wql,
    basicstyle=\small\ttfamily,
    keywordstyle=\color{darkBlue},
    #1
}}{}

\lstdefinelanguage{neo4j}{
  language=SQL,
  morekeywords=[1]{RETURN, as, WITH},
  keywordstyle={[1]\color{darkBlue}\ttfamily},
  deletekeywords={local},
  stringstyle=\ttfamily\color{orange},
  showstringspaces=false,
  comment=[l]{//},
  morecomment=[s]{/*}{*/},
  commentstyle=\color{commentgreen}\ttfamily,
  numberstyle=\scriptsize\color{gray},
  rulecolor=\color{gray},
  tabsize=2,
}

\lstnewenvironment{neo4j}[1][]{\lstset{
    language=neo4j,
    basicstyle=\small\ttfamily,
    keywordstyle=\color{darkBlue},
    #1
}}{}

\lstdefinelanguage{datalog}{
    language=c++,
    deletekeywords={NAME},
    morekeywords=[1]{decl, unsigned, symbol},
    keywordstyle={[1]\color{darkBlue}\bfseries\ttfamily},
    commentstyle=\color{commentgreen}\ttfamily,
    stringstyle=\color{orange},
    numberstyle=\scriptsize\color{gray},
    rulecolor=\color{gray},
}
\lstnewenvironment{datalog}[1][]{\lstset{
    language=datalog,
    basicstyle=\small\ttfamily,
    keywordstyle=\color{darkBlue},
    #1
}}{}

\usepackage{array}
\usepackage{tabularx}
\usepackage{booktabs}
\usepackage{arydshln}

\usepackage{multirow}

\newcolumntype{L}[1]{>{\raggedright\let\newline\\\arraybackslash\hspace{0pt}}m{#1}}
\newcolumntype{C}[1]{>{\centering\let\newline\\\arraybackslash\hspace{0pt}}m{#1}}
\newcolumntype{R}[1]{>{\raggedleft\let\newline\\\arraybackslash\hspace{0pt}}m{#1}}

\usepackage{soul}

\usepackage{wasysym}

\usepackage{colortbl}
\usepackage{diagbox}
\usepackage{bigstrut}

\begin{document}

\title{Wasmati: An Efficient Static Vulnerability Scanner for WebAssembly}

\author{Tiago Brito$^*$}
\email{tiago.de.oliveira.brito@tecnico.ulisboa.pt}
\affiliation{%
  \institution{INESC-ID / IST, Universidade de Lisboa, Portugal}
  $^*$Corresponding Author
}

\author{Pedro Lopes}
\email{pedro.daniel.l@tecnico.ulisboa.pt}
\affiliation{%
  \institution{INESC-ID / IST, Universidade de Lisboa, Portugal}
}

\author{Nuno Santos}
\email{nuno.m.santos@tecnico.ulisboa.pt}
\affiliation{%
  \institution{INESC-ID / IST, Universidade de Lisboa, Portugal}
}

\author{José Fragoso Santos}
\email{jose.fragoso@tecnico.ulisboa.pt}
\affiliation{%
  \institution{INESC-ID / IST, Universidade de Lisboa, Portugal}
}

\begin{abstract}
WebAssembly is a new binary instruction format that allows targeted compiled code written in high-level languages to be executed with near-native speed by the browser's JavaScript engine.
However, given that WebAssembly binaries can be compiled from unsafe languages like C/C++, classical code vulnerabilities such as buffer overflows or format strings can be transferred over from the original programs down to the cross-compiled binaries. As a result, this possibility of incorporating vulnerabilities in WebAssembly modules has widened the attack surface of modern web applications.

This paper presents Wasmati, a static analysis tool for finding security vulnerabilities in WebAssembly binaries. It is based on the generation of a code property graph (CPG), a program representation previously adopted for detecting vulnerabilities in various languages but hitherto unapplied to WebAssembly. We formalize the definition of CPG for WebAssembly, introduce techniques to generate CPG for complex WebAssembly, and present four different query specification languages for finding vulnerabilities by traversing a program's CPG. We implemented ten queries capturing different vulnerability types and extensively tested Wasmati on four heterogeneous datasets. We show that Wasmati can scale the generation of CPGs for large real-world applications and can efficiently find vulnerabilities for all our query types. We have also tested our tool on WebAssembly binaries collected in the wild and identified several potential vulnerabilities, some of which we have manually confirmed to exist unless the enclosing application properly sanitizes the interaction with such affected binaries.

\end{abstract}


\maketitle


\newcommand{\wasm}{WebAssembly\xspace}

\section{Introduction}

WebAssembly~\cite{wasm} is an emerging binary code format designed for speeding up the Web. Currently supported by all major browsers, WebAssembly bytecode runs on a stack-based virtual machine that takes advantage of the local hardware capabilities to achieve near-native execution performance. Also known as Wasm, WebAssembly is a low-level assembly-like language as well as a compilation target for higher-level programming languages like C++. This feature has prompted swathes of pre-existing C++ libraries and applications to be ported to run in the browser~\cite{awesomewasm}, boosting the adoption of Wasm in the Web~\cite{crypto}. WebAssembly's portability and efficiency have led to its usage far beyond the browser, being employed for running sandboxed code of server-side web applications~\cite{nodejs,wasmer,wasi}, IoT apps~\cite{wasmiot}, edge computing logic~\cite{lucet}, or smart contracts~\cite{smartcontracts}.

However, WebAssembly opens up new avenues for the introduction of security vulnerabilities in the Web and other Wasm-based environments~\cite{lehmann2020everything}. Albeit the extensive safety mechanisms incorporated into WebAssembly itself, many coding errors and unsafe functions written in C or C++ can still be transposed into WebAssembly binaries. As a result, web application code that depends on WebAssembly modules may now become crippled by the introduction of classic forms of software security flaws, such as format strings, use-after-free, double free, or buffer overflow vulnerabilities. Preliminary studies have shown that when such flaws are present in Web applications, they may be leveraged for launching several web attacks, such as cross-site scripting (XSS) or code injections~\cite{blackhat}.

Given the pace at which many software projects are being ported to WebAssembly and being adopted worldwide~\cite{hilbig2021empirical}, we foresee the imminent danger of pre-existing vulnerabilities to creep into many applications and potentially cause much damage in the near future. As a preemptive measure to tackle this risk, our focus in this work is to help eliminate potential security flaws in WebAssembly binaries, regardless of the specific target application where these binaries are used, e.g., as part of a client-side web page running alongside JavaScript code, as a server-side module running on a Node.js application, or even as a component of a browser extension.

To root out many security flaws from Wasm modules, one can compile these modules from code written in safe programming languages like Go or Rust. However, porting countless software written in C/C++ to a safe language would be a daunting and impractical endeavor. Alternatively, checking and fixing vulnerabilities in C/C++ code using vulnerability scanner tools would proactively counter the transposition of security bugs to Wasm binaries. However, WebAssembly applications are normally composed of multiple libraries or modules, most of which have been precompiled independently by third parties. Access to the original sources may not even be possible (e.g., proprietary code). We then propose a third alternative based on the direct analysis of WebAssembly binaries. 

This paper presents Wasmati, an efficient static analysis tool for finding vulnerabilities in Wasm binaries. Wasmati can be shipped in the form of a library that can be linked to other programs (e.g., infrastructure-level software or security analysis frameworks) or packaged as a standalone CLI program. It can then be used for checking vulnerabilities at the development stage (e.g., in the software development toolchain) or in the production stage (e.g., for analyzing client-side web applications in the wild). Currently, our tool can analyze Wasm binaries generated by the popular Emscripten~\cite{emscripten} compiler. Wasmati is parameterized by a collection of vulnerability queries. By enriching this set of queries, one can augment the typology of vulnerabilities that Wasmati can detect.

To build Wasmati, we adopt a recently proposed static program analysis technique named {\em code property graph} (CPG)~\cite{cpg,yamaguchi2015automatic}. CPGs have proved to be powerful constructs for building vulnerability scanners at the source code level for languages such as C/C++~\cite{shiftleftSite,joern,codeql}, Java~\cite{shiftleftSite,codeql}, Python~\cite{shiftleftSite,codeql}, or PHP~\cite{backes2017efficient}, and also at the low virtual machine code level for LLVM~\cite{shiftleftSite} and Java bytecode~\cite{plume}. Wasmati is the first tool of this kind to analyze WebAssembly bytecode.

In the design of Wasmati, the low-level nature and specific semantics of WebAssembly brought about several non-trivial challenges. In the construction of the CPG, a major obstacle was in devising a scalable technique to analyze complex Wasm binaries. In Wasm code, the number of nodes of a CPG grows dramatically. Adding to the fact that CPG requires multiple graph traversals for adding edges relative to its distinct subgraphs, the overall construction of a CPG can become prohibitively slow even for modest-sized libraries. For query specification, a core challenge is to represent CPG search patterns while satisfying three requirements: i) expressiveness power, i.e., allow the detection of vulnerability types through the traversal and inspection of the CPG's properties, ii) conciseness and simplicity, i.e., allows security analysts to easily specify new queries, and iii) performance,  i.e., the evaluation of queries must terminate in an acceptable time, viz. in a few seconds or minutes.

Wasmati features an optimized CPG data structure and generator engine that significantly speeds up the CPG generation process. Some of the optimization strategies include i) enriching the GPG graph with additional annotations, ii) caching intermediate results, and iii) employing efficient graph traversal algorithms. As for query specification, to strike a good balance between expressiveness power, conciseness, and performance, we developed a domain-specific language named Wasmati Query Language (WQL). WQL allows security analysts to specify vulnerability patterns in a developer-friendly way while offering efficient query execution times. To study how WQL performs in comparison with alternative query specification languages, Wasmati features a generic CPG query engine that enables queries to be specified and executed in three additional languages: C++, Datalog, and Neo4j's Cypher.

We implemented 10 different vulnerability queries and extensively evaluated Wasmati using four different Wasm binary datasets. Wasmati found 100 vulnerabilities out of 108 present in these datasets, representing a precision of 92.6\%. This shows that vulnerabilities in WebAssembly can be modeled using CPGs. Wasmati's generation of CPGs can scale for large real-world applications. Using SPEC CPU 2017, the construction time averaged 58 seconds per binary. Graph construction is polynomial in time given the graph's size. We executed the 10 queries in generated CPGs of SPEC CPU 2017 with an average execution time of 77 seconds. We tested Wasmati on a dataset comprising real-world WebAssembly binaries and found several potential vulnerabilities. We manually analyzed some of the flagged vulnerabilities and confirmed that they can be triggered by crafted inputs provided into the affected module.

In summary, this paper makes the following contributions: (1) first formalization of CPGs for WebAssembly, (2) techniques for efficient generation of WebAssembly CPG and specification of vulnerability queries for Wasm code, (3) robust implementation of the Wasmati tool, (4) extensive evaluation using real-world Wasm binaries and queries written in four different languages.

\section{Background and Overview}
\label{sec:background}

In this section, we introduce WebAssembly, motivate our work using a real-world vulnerability that persists in WebAssembly, provide an overview of how Wasmati detects this vulnerability and, finally, clarify the design goals and scope of our new tool.

\subsection{Background on WebAssembly}
\begin{figure}[t]
    \centering
    \includegraphics[width=\columnwidth]{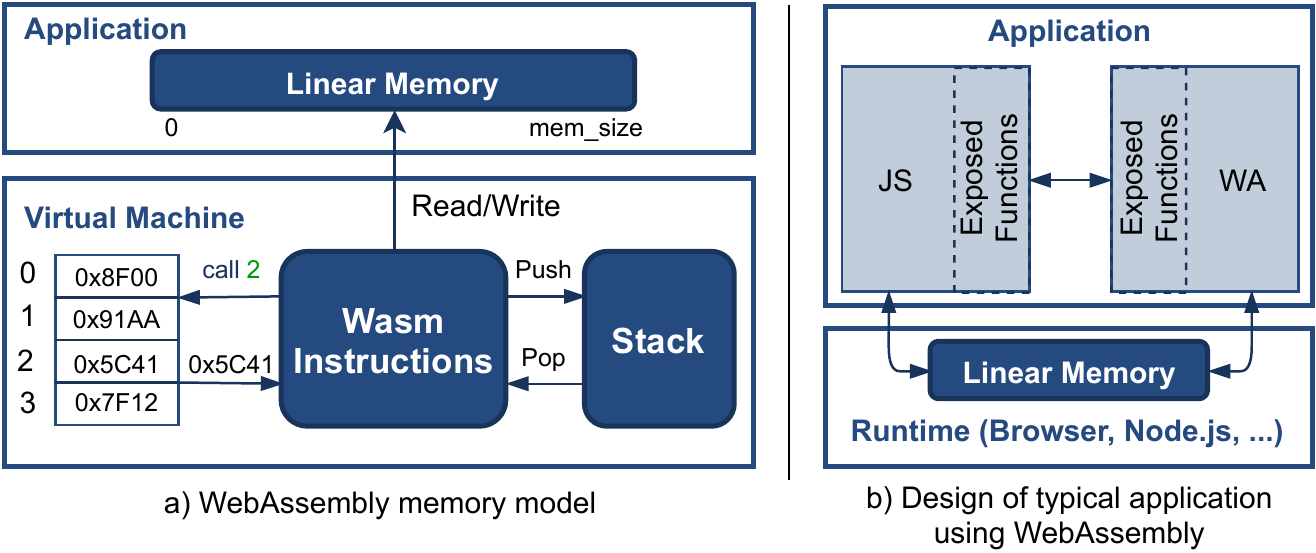}
    \caption{\small WebAssembly high-level architecture.}
    \label{fig:wasm_models}
    \vspace{-0.4cm}
\end{figure}

The WebAssembly binary is a sequence of instructions that are executed on a stack-based machine.
Simple instructions perform operations on data; they consume their operands from the stack and produce a result that is placed on the stack. Control instructions alter the control flow of the code. A program can call functions directly or indirectly through a function table. Figure~\ref{fig:wasm_models} a) shows a direct call of the index 2. Indirect calls allow the emulation of function pointers and polymorphism in OOP languages such as C++. The table index value is pushed into the stack, evaluated at execution time, and the function indexed by that value is executed. Function tables can be defined in modules. A module represents the binary format of Wasm that has been compiled.

WebAssembly has only four primitive types: \verb|i32|, \verb|i64|, \verb|f32| and \verb|f64|. The first two represent integers with 32 and 64 bits respectively, whereas the last two denote 32 and 64-bit floating-point data. Global variables, local variables, and return addresses are managed in the stack. All non-scalar types, such as strings, arrays, and other buffers, must be stored in \textit{linear memory}, which is a contiguous, untyped, byte-addressable array, multiple of 64Kib. A program can load/store values from/to linear memory at any byte address. A trap occurs if an access is not within the bounds of the current memory size.

WebAssembly binaries are executed by a runtime engine. They are normally deployed in the form of modules that pertain to a larger JavaScript application (see Figure~\ref{fig:wasm_models} b)). Applications use dedicated JavaScript code to bootstrap the Wasm module into a sandboxed environment and to interface with external resources (e.g. DOM). JavaScript code and WebAssembly modules can mutually expose function calls and communicate with each other through shared linear memory. WebAssembly-based applications can run on various platforms, most commonly on the browser, but also on web servers (e.g., powered by Node.js) or desktops.

\subsection{Practical Vulnerability Example}

\begin{figure}[t]
	\centering
	\small
		\begin{cpp}[numbers=left, frame=l]
void get_token(FILE *pnm_file, char *token) {
  int i = 0;
  int ret;
  // (...)
  do {
    ret = fgetc(pnm_file);
    if (ret == EOF) break;
    i++;
    token[i] = (unsigned char) ret;
  } while ((token[i] != '\n') && (token[i] != '\r')
        && (token[i] != ' '));
  token[i] = '\0';
  return;
}
\end{cpp}
\vspace{-0.5cm}
\caption{\small Buffer overflow in libpng (CVE-2018-14550).}
\label{code:libpng}
\vspace{-0.3cm}
\end{figure}

\begin{figure}[t]
	\centering
	\small
		\begin{cpp}[numbers=left, frame=l]
void main() {
  std::string img_tag = "<img src='data:image/png;base64,";
  // bad input: AAAA...AA<script>alert("XSS!")</script><!--
  pnm2png("input.pnm", "output.png");  // CVE-2018-14550
  img_tag += file_to_base64("output.png") + "'>";
  emcc::global("document").call("write", img_tag);
}
\end{cpp}
\vspace{-0.5cm}
\caption{\small Exploit for CVE-2018-14550 by Lehman et al.~\cite{lehmann2020everything}.}
\label{code:libpng-exploit}
\vspace{-0.3cm}
\end{figure}

Figure~\ref{code:libpng} presents an example of a real stack buffer overflow vulnerability (CVE-2018-14550~\cite{CVE-2018-14550}) existing inside libpng, which is the official PNG reference library and it is widely used by many applications. Affecting the function \texttt{get\_token}, this vulnerability can persist when this code is compiled from C to WebAssembly and it has been previously used by Lehman et al.~\cite{lehmann2020everything} to showcase how vulnerable code written in memory-unsafe languages can be transferred to WebAssembly modules. By developing the exploit presented in Figure~\ref{code:libpng-exploit}, the same authors have shown that this flaw can be harnessed to launch a cross-site scripting (XSS) attack. We leverage the same example to motivate our work and then, in the next section, we illustrate how our tool can detect this vulnerability.

The buffer overflow vulnerability shown in Figure~\ref{code:libpng} can be triggered when converting a PNM file to a PNG file using libpng. This operation calls \texttt{get\_token} by providing a 16-byte length local buffer as the \texttt{token} parameter. Inside \texttt{get\_token} no check is performed to assess if this buffer is being written beyond its 16-byte length, which allows for a stack buffer overflow to occur whenever the \texttt{pnm\_file} parameter exceeds 16 bytes. Normally, when libpng is compiled to native binary code, stack canaries prevent this vulnerability from being exploited. Even if stack canaries are not employed by the compiler, the buffer overflow is limited to the stack. However, in WebAssembly this vulnerability can be freely exploited without any of these mitigation strategies. Additionally, when taking into account different linear memory layouts from WebAssembly compilers and backends, this vulnerability can lead to writes not only in the stack, but also the heap and data sections of the memory.

The C++ code in Figure~\ref{code:libpng-exploit} represents a simplified version of a service that converts images using the libpng library and then displays the converted image by writing the content to the DOM using the \texttt{document.write} function. This code converts a PNM image to PNG (line 4), encodes the image content in base64, appends the image content into the \texttt{img} tag (line 5), and then adds the tag into the document by manipulating the DOM (line 6). Since the image content is embedded into the DOM as a base64-encoded string, it normally cannot lead to XSS. However, the stack-based buffer overflow in libpng allows the attacker to overwrite higher addresses, including the heap, which holds the C++ string with the \texttt{img} tag (line 2). This way, the attacker can use a crafted malicious input, such as the \texttt{script} tag string containing an alert (line 3) as the content of the image to convert and override the \texttt{img} tag with the new crafted \texttt{script} tag, thus causing an XSS attack.

\begin{figure}
    \centering
    \footnotesize
    \begin{webassembly}[numbers=left, frame=l, framesep=1em]
(module
  (func $get_token (param $pnm_file i32)(param $token i32)
  (local $i i32)
  (local $ret i32)
  ;; (...)
  loop $L4
      block $B5
        local.get $pnm_file
        call $fgetc
        local.tee $ret
        i32.const -1
        i32.eq ;; ret == EOF
        br_if $B5
        local.get $token
        local.get $i
        i32.const 1
        i32.add ;; i++
        local.tee $i
        i32.add ;; token + i
        local.get $ret
        i32.store8 ;; token[i] = ret
        local.get $ret
        i32.const 10
        i32.eq ;; token[i] == '\n'
        br_if $B5
        local.get $ret
        i32.const 13
        i32.eq ;; token[i] == '\r'
        br_if $B5
        local.get $ret
        i32.const 32
        i32.ne ;; token[i] == '\r'
        br_if $L4
    end
  end
  ;; (...)
  i32.const 0))
    \end{webassembly}
    \vspace{-0.3cm}
    \caption{\small WebAssembly textual representation of the buffer overflow vulnerability in libpng (CWE-2018-14550)}
    \label{fig:libpng-wat}
    \vspace{-0.3cm}
\end{figure}

\subsection{Finding Vulnerabilities With Wasmati}

\begin{figure*}[t]
    \centering
    \includegraphics[width=\textwidth]{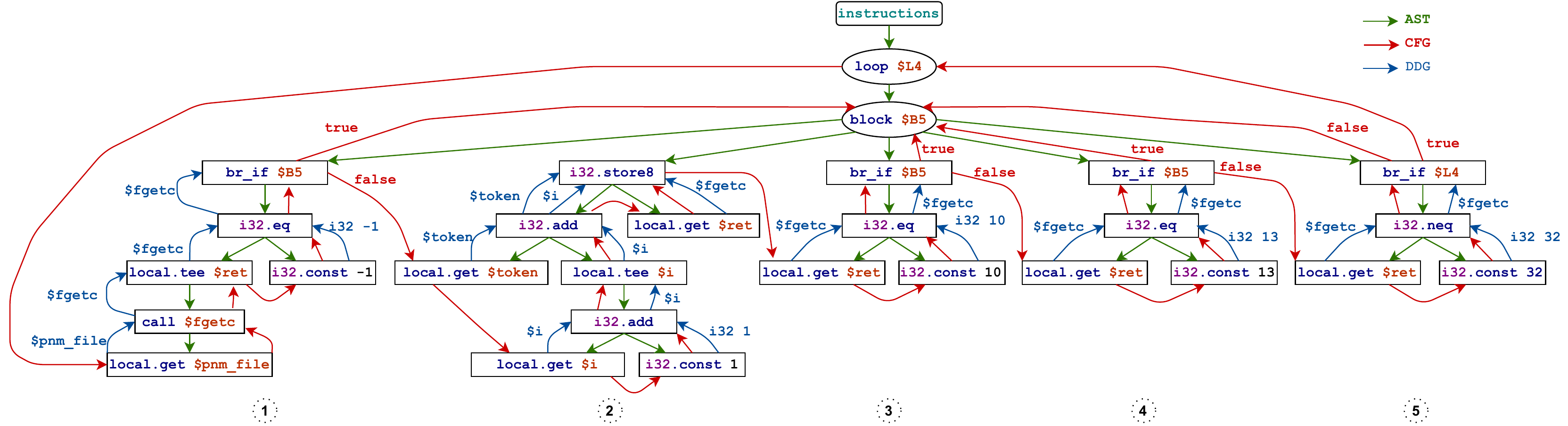}
    \caption{\small Simplified Code Property Graph generated by Wasmati for WebAssembly code fragment depicted in Figure~\ref{fig:libpng-wat}.}
    \label{fig:wasmati_cpg}
    \vspace{-0.4cm}
\end{figure*}

To cater for the static detection of C/C++-style vulnerabilities in Wasm code, we generate and analyze WebAssembly-specific Code Property Graphs (CPGs). GPGs are graph-based data structures which include information about the analyzed code in the form of property-value pairs (hence the name CPG). These pairs can then be queried in different ways to search for different types of vulnerabilities that can exist in the analyzed code. Using our CPG specification for WebAssembly, Wasmati can detect the buffer overflow vulnerability in the libpng presented in Figure~\ref{code:libpng} by analyzing the corresponding Wasm code representation listed in Figure~\ref{fig:libpng-wat}. Next, we give an overview of Wasmati CPGs and then explain how this vulnerability can be detected by searching for a specific pattern in the CPG's sub-graphs of the corresponding \wasm code.

\mypara{1. Generating the CPG:} Figure~\ref{fig:wasmati_cpg} depicts part of the CPG generated by Wasmati for the code shown in Figure~\ref{fig:libpng-wat}. The WebAssembly CPG is comprised of four different graps: Abstract Syntax Tree (AST), Control Flow Graph (CFG), Data Dependency Graph (DDG), and Call Graph (CG). The first three are portrayed in Figure~\ref{fig:wasmati_cpg} being distinguished by the edge colors. For clarity, AST edges are illustrated by green edges and node properties are not represented in the figure. A full description of node properties can be found in the appendix. The CFG explicitly describes the order by which instructions are executed and the conditions necessary for taking a particular execution path. The CFG edge, illustrated in red, may contain a label that helps identify the condition that allows that particular flow to be followed. The DDG explicitly represents dependencies between the instructions of the program. Dependencies can be function calls, local variables, global variables or constants. Each dependency is expressed as an edge painted in blue and its label refers to the name of the dependency. The CG is essential to support inter-procedural analysis and is a simple directed graph that connects call nodes and the root of the corresponding called function. It is not visible in Figure~\ref{fig:wasmati_cpg} as we are not analyzing an inter-procedural case, but we explain its importance in Section~\ref{sec:cpg}. To better visualize the CPG, the nodes in this figure are arranged in a tree layout (corresponding to the AST) with five sub-trees which are numbered from 1 to 5. Each sub-tree can be seen as a bigger statement which is a composition of different instructions. This layout and an understanding of the CPG helps us go through the steps needed to find the vulnerability reported in CVE-2018-14550.

\mypara{2. Querying the CPG:} The idea to find this type of vulnerability -- i.e., buffer overflows -- in WebAssembly code is to analyze the CPG searching for characteristic patterns. In WebAssembly, buffer overflows happen when a buffer is incorrectly used inside a loop, e.g., when handling strings. A vulnerability may occur because the index of the buffer is incremented and the buffer's bounds are not being checked as part of the loop's exit condition. The idea to detect buffer overflows then is to search for instructions in a loop where the AST descendants (loop's block and condition): 1) contain a local variable \verb|$i| representing the index that is being incremented, 2) have a store instruction that depends on a buffer and \verb|$i| (assignment), and 3) lack an exit \verb|br_if| whose condition test verifies the boundaries of \verb|$i|. Thus, to find the vulnerability in our running example, we can query the CPG looking for patterns that satisfy these three conditions.
For the first, we look for instructions \verb|i32.add| that have incoming local DDG edge for \verb|$i| and a constant DDG edge (which is the increment value).
For the second, we simply look for \verb|store| instruction with incoming local DDG edge for \verb|$i|.
And lastly, we search for \verb|br_if| instructions and query its AST descendants (representing the composite condition) for the existence of comparison instruction with incoming local DDG edge for \verb|$i|. This idea can then be generalized to look for other types of vulnerabilities by adjusting the CPG query accordingly.

\subsection{Design Goals and Scope}

Our main goal is then to build a static analysis tool that can detect security flaws that can be propagated from the original programs (typically written in a high-level language like C/C++) into WebAssembly binaries.
We are also interested in building efficient query engines that can help us study several inherent trade-offs between query expressiveness power, conciseness, and performance.

The analysis implemented by our tool will be focused on individual Wasm modules independently of how they are used by a given application or the platform where they are deployed. To examine how a particular security flaw in a Wasm module would manifest itself in a full-blown application, it would be necessary to analyze not only individual WebAssembly binaries but also all other application components (including JavaScript code), which fall outside the scope of this paper due to a significant added complexity.

\section{Wasmati Architecture}
\label{sec:architecture}

\begin{figure}[t]
    \centering
    \includegraphics[width=0.95\columnwidth]{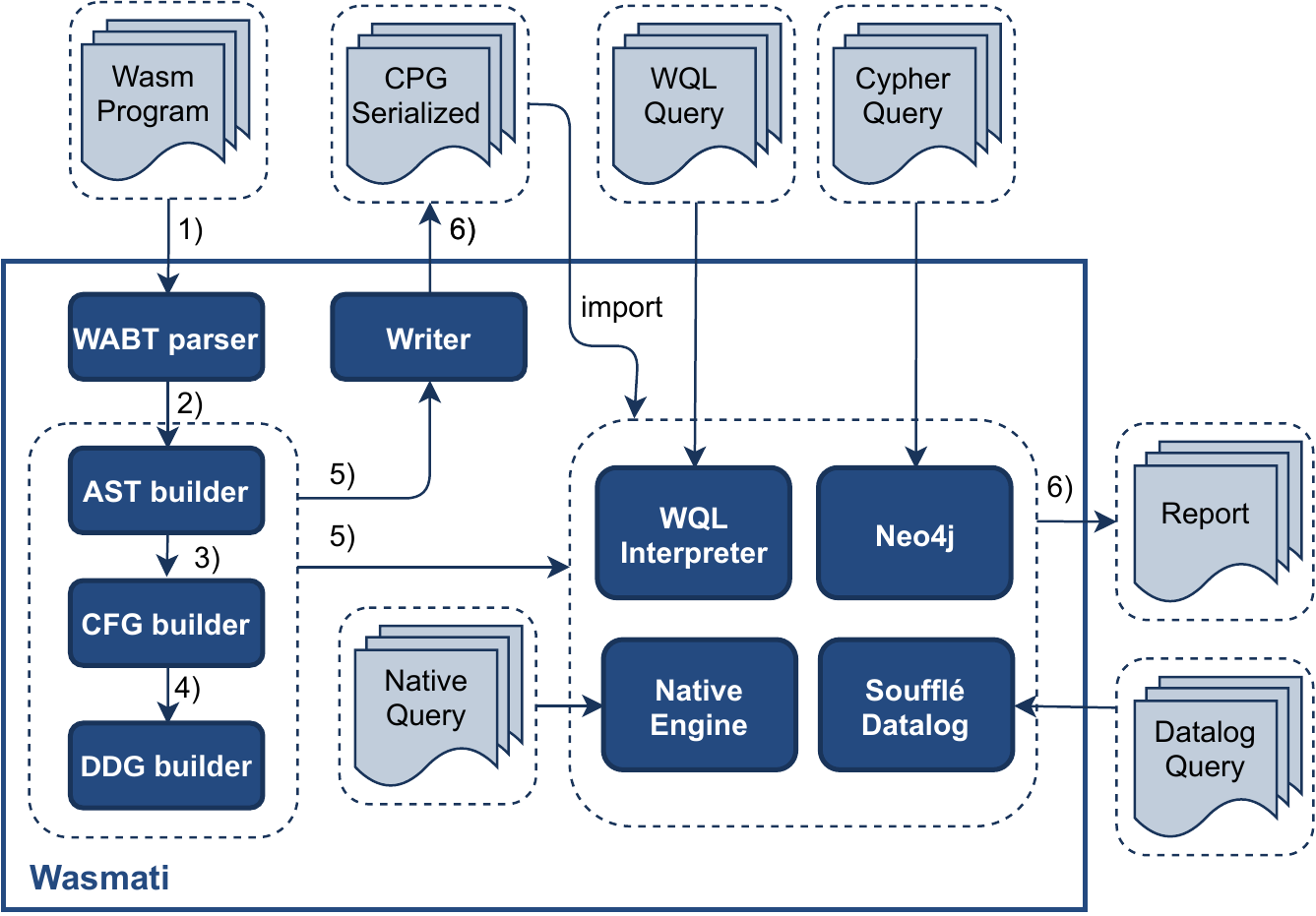}
    \vspace{-0.1cm}
    \caption{\small Wasmati architecture.}
    \label{fig:wasmati_arch}
    \vspace{-0.3cm}
\end{figure}

We present Wasmati, a new static analysis tool based on the generation of CPGs for finding security vulnerabilities in WebAssembly binaries. We implemented Wasmati using C++11 in about 12.350 lines of code. Figure~\ref{fig:wasmati_arch} represents the internal components of our tool. Wasmati consists of two processing pipelines: the \textit{CPG generator pipeline} and the \textit{query engine pipeline}. The former is responsible for analyzing an input Wasm program and generating the corresponding CPG data structure into a file. The query engine pipeline loads the CPG from this file and executes a series of queries by searching for specific vulnerability patterns in the CPG. Below, we briefly describe the inner workings of each processing pipeline.

\mypara{CPG generator pipeline:} Wasmati receives a Wasm module, in binary or text format, which will be fed into WebAssembly Binary Toolkit (WABT)~\cite{wabt}. We used WABT, an open-source project maintained by the official WebAssembly community, to parse the WebAssembly's binary/text formats. The parser produces a list of functions. This list is processed by a chain of components which progressively build the CPG. First, this list is provided as input to the abstract syntax tree (AST) builder which creates the AST. Then, the control flow graph (CFG) builder generates the CFG by making another iteration over the function list; the call graph (CG) is also generated in this stage. Lastly, the data dependency graph (DDG) builder creates the DDG. The resulting CPG is then written to a file in one of several possible serialization formats.

\mypara{Query engine pipeline:} The query pipeline supports four query languages/back-ends to execute the CPG traversals:

\noindent \textit{1. WQL:} is the Wasmati Query Language, a DSL that eases the writing of CPG traversals for WebAssembly. The queries written in WQL are interpreted by the WQL interpreter. WQL strikes a good balance between expressiveness power, specification simplicity, and performance. The following back-ends explore different trade-offs in the design space which we explain more thoroughly in Section~\ref{sec:queries}.

\noindent \textit{2. Native:} consists of a query API that enables Wasmati to be extended with additional queries written in C++. Adding new queries requires the re-compilation of Wasmati.

\noindent \textit{3. Neo4j:} is a graph database that can import the serialized CPG and be traversed using Cypher Query Language (CQL). Wasmati comes with a Dockerfile and scripts to import and run queries automatically. A new query can be added to a specific folder.

\noindent \textit{4. Datalog:} is a declarative logic programming language that allows the Wasm CPG to be queried in a deductive manner. Wasmati uses Soufflé's~\cite{jordan2016souffle} flavoured datalog and its engine. Alongside a Docker configuration file, which automatically imports and executes datalog queries, Wasmati also comes with a library containing common predicates and definitions to query the CPG.

\mypara{Currently implemented queries:}  
We focused on five main classes of common vulnerabilities in C/C++ that can still persist in the compiled Wasm code: format strings, dangerous functions, use-after-free/double-free and different variations of tainted-style vulnerabilities and buffer overflows. To detect these types of vulnerabilities, we implemented a total of ten different queries in each of Wasmati's supported query languages.
Regarding tainted-style vulnerabilities, we implemented three variants: tainted call indirect, tainted function-to-function, and tainted local-to-function.
In the first, we query the taintability of the last argument of \verb|call_indirect| which controls the function to be called.
The second looks for the classical result of an input source reaching a sink. The third looks for a possible tainted parameter that reaches a sink.
With respect to buffer overflows, we also target three variants: static buffer overflow, where we compare the size of the buffer against the size of the data being written to it; 
malloc buffer overflow, similar to static buffer overflow but the buffer is allocated dynamically with a constant value; and loop buffer overflow, where we search for buffer writes without boundary checks -- this variant can detect a buffer overflow in libpng (CVE-2018-14550)~\cite{CVE-2018-14550}.

In the following sections, we describe in detail the core challenges and operations performed by the processing pipelines of Wasmati, namely building and querying WebAssembly CPGs.

\section{Building WebAssembly CPGs}
\label{sec:cpg}

\newcommand{\props}{\mathcal{K}}
\newcommand{\prop}{\kappa}
\newcommand{\vals}{\mathcal{V}}
\newcommand{\val}{v}
\newcommand{\jose}[1]{{\color{blue}#1}}
\newcommand{\wasabi}[1]{\textsc{Wasabi}}

\newcommand{\constdep}[2]{\mathtt{C}_{#1}(#2)}
\newcommand{\localdep}[2]{\mathtt{LV}_{#1}(#2)}
\newcommand{\globdep}[2]{\mathtt{GV}_{#1}(#2)}
\newcommand{\funcdep}[2]{\mathtt{F}_{#1}(#2)}
\newcommand{\id}[0]{id}
\newcommand{\sint}[0]{{\sf i32}}

\newcommand{\dep}[0]{\varphi}
\newcommand{\deps}[0]{\Phi}
\newcommand{\depset}[0]{\phi}
\newcommand{\globvars}[0]{\mathcal{GV}}
\newcommand{\lvars}[0]{\mathcal{LV}}
\newcommand{\gsto}[0]{\hat{g}}
\newcommand{\lsto}[0]{\hat{l}}
\newcommand{\stack}[0]{\hat{st}}

\newcommand{\transfer}[0]{\mathcal{T}}
\newcommand{\insts}[0]{\mathcal{I}}
\newcommand{\astates}[0]{\hat{\mathcal{S}}}
\newcommand{\tvar}[1]{\mathtt{#1}}

\newcommand{\transrule}[4]{\mathcal{T}(\mathtt{#1}, #2) \hookrightarrow_{#4} #3}

 \let\smallish\small
  
  \newcommand{\hbra}{
  \hbox to \linewidth{\vrule width0.3mm height 1.8mm depth-0.3mm
    \leaders\hrule height1.8mm depth-1.5mm\hfill
    \vrule width0.3mm height 1.8mm depth-0.3mm}}
\newcommand{\hket}{
  \hbox to \linewidth{\vrule width0.3mm height1.5mm
    \leaders\hrule height0.3mm\hfill
    \vrule width0.3mm height1.5mm}}
  
\newcommand{\ratio}{.375}
\newenvironment{display}[1]{\vspace{-0.5ex}\smallish
\begin{tabbing}
    \hspace{1.5em} \= \hspace{\ratio\linewidth-1.5em} \= \hspace{1.5em} \= \kill
    \textbf{#1}\\[-.8ex]
    \hbra\\[-.8ex]
  }{\\[-.8ex]\hket
  \end{tabbing}\vspace{-1ex}}

The specific features of \wasm introduce non-trivial obstacles to the construction of \wasm CPGs which required us to: (i) formalize a tailor-made data structure for representing \wasm CPGs, (ii) develop a specific data flow analysis algorithm for computing \wasm DDGs, and (iii) incorporate a set of optimizations to speed up the process of building CPGs. In this section, we present how Wasmati builds \wasm CPGs emphasizing these main distinguishing features of our system.

\subsection{Specification of \wasm CPGs}

Formally, a CPG $G = (V, E, \mu)$ is a triple such that: (i) $V$ is the set of nodes, (ii) $E$ is the set of edges connecting the nodes in $V$, and (iii) $\mu : V \cup E \rightarrow \props \rightharpoonup \vals$ is a total function linking each vertex/edge to a (partial) map connecting its properties to their values. For instance, let $n$ be a node in $V$, the function $\mu(n) : \props \rightharpoonup \vals$ maps the properties of $n$ to their corresponding values. For clarity, we use $\mu(n, \prop)$ instead of $\mu(n)(\prop)$ to denote the value of property $\prop$ of node $n$.
Not all nodes/edges define all the properties in $\props$. More concretely, each type of node/edge defines its own specific properties; for instance, store/load instruction nodes define a property \emph{offset} for holding the offset associated with their respective~instructions. 

A CPG $G = (V, E, \mu)$ can be decomposed into four sub-graphs, respectively corresponding to the AST, CFG, CG, and DDG of the program to be analyzed. For instance, we write $G_{AST} = (V, E_{AST}, \mu_{AST})$ for the AST component of $G$. While these four graphs share the same underlying set of nodes, $V$, they have different edges and they store different property-value pairs. Importantly, each edge $e \in E$ is associated with a property \emph{type}, indicating the sub-graph to which it belongs (i.e. $\mu(e, \textit{type}) = \textsc{AST}$ means that $e \in E_{AST}$).

\mypara{Abstract Syntax Tree (AST):}
To build the AST, we leverage the official open-source parser included in WABT\footnote{\url{https://github.com/WebAssembly/wabt}}. However, the AST produced by the WABT parser is a flat tree, and for the purpose of CPG query traversal this lack of structure makes it hard to analyze instructions that take multiple input arguments (e.g., \texttt{i32.add}, or function calls). To overcome this limitation, we re-arrange the array of instructions produced by the parser to make sure that direct dependencies are taken into account resulting in a hierarchical organization. For instance, in the program of Figure~\ref{fig:wasmati_cpg}, the instructions \verb|local.get $i| and \verb|i32.const 1| become children of \verb|i32.add| (sub-tree 2), while the original parser represents the three instructions at the same level. To perform this re-organization, Wasmati implements an ``AST folding'' algorithm (shown in the appendix). Our AST includes additional nodes that store meta-information concerning the WebAssembly program; for instance, we use the node \texttt{return} to pinpoint the expression that computes the return value of a function.

\mypara{Control Flow Graph (CFG):}
In our case, building the CFG is relatively simple because the \wasm control flow is structured and can be statically verified. In contrast to typical native binaries (e.g., for x86 or Arm), in \wasm there are no relative jumps and a jump target cannot be an instruction from the middle of a block. Wrong paths are not allowed and are validated before execution by the runtime. As a result, the \wasm CFG is mostly linear: each instruction has a CFG edge connecting it to the next instruction in the code. The exceptions are the branching instructions \verb|if|, \verb|br_if| and \verb|br_table|, which have more than one successor node. Unsurprisingly, the two outgoing edges of \verb|if| and \verb|br_if| are labeled with either \textbf{true} or \textbf{false} to distinguish the \emph{then} branch from the \emph{else} branch. Analogously, the instruction \verb|br_table|, which works as a \verb|switch| statement in a high-level language, has its outgoing branches annotated with the concrete values that cause the control to be transferred to each of its branches. 
Finally, branching labels are stored in the property \emph{label} of the corresponding CFG edge. For instance, given a CFG edge $e \in E_{CFG}$, $\mu(e, label) = true$ means that $e$ is the \emph{then} branch of a \verb|br_if| instruction.

\mypara{Call Graph (CG):}
Call graphs are essential to support inter-proce-dural analysis. In a nutshell, a call graph is a simple directed graph that connects call nodes (i.e., nodes representing call instructions) to the root nodes of the corresponding functions. To implement a CG in \wasm, we have to consider both \emph{direct calls} and \emph{indirect calls}. Direct calls are processed straightforwardly: each direct call instruction is directly connected to the node representing the function being called. However, indirect calls are more difficult to analyze, as the index of the function being called is computed at runtime. An indirect call is a mechanism that allows polymorphism from OOP source languages (e.g. C++) and is dependent on its execution. As a result, it is impossible to determine statically which functions will be executed. In fact, multiple functions can be executed depending on different executions.

We solve this challenge as follows. In \wasm, for a dynamic call to be executed successfully: 1) the called function must be stored in the function table, and 2) the signature of the called function must coincide with the signature supplied to the call instruction. Hence, for indirect calls, we simply connect every indirect function call to all the function nodes whose signatures match the statically supplied signature. Finally, all CG edges have a single property \emph{type} with value \textsc{CG}, tagging them as part of the call graph. 

\mypara{Data Dependency Graph (DDG):}
A DDG explicitly represents dependencies between the instructions of the program to be analyzed. In the original CPG paper~\cite{cpg}, it is comprised of both \textit{data dependencies} and \textit{control dependencies}, which the authors coalesce into a Program Dependency Graph (PDG). An instruction $inst_2$ data-depends on another instruction $inst_1$, if $inst_2$ uses a variable defined by $inst_1$. In turn, $inst_2$ control-depends on $inst_1$, if the execution of $inst_2$ depends on $inst_1$ (e.g., $inst_1$is a branch instruction).

However, the original PGD definition for CPGs is not ideal for \wasm. For one, keeping track of both dependency types leads to inefficiencies and scalability bottlenecks when constructing and when querying the graph. Compiled \wasm code contains many instructions organized in long chains of conditional blocks and loops which may result in an exceedingly large number of control dependencies edges (in our preliminary experiments, comprising in some cases 98\% of the total edges of a program's CPG and bloating its size up to 65 times). Secondly, the original PDG definition~\cite{cpg} is too coarse. In \wasm, we require richer semantics that allows us to reason about different variable scopes (local and global), return values from function calls, and constant value propagation. For instance, we need to differentiate variables from constants given that oftentimes the compilation of an instruction in C that uses a constant value translates into multiple \wasm instructions accessing local variables, making it difficult to keep track of the constant in the \wasm binary. 

\begin{figure}
    \centering
    \begin{subfigure}[b]{0.32\columnwidth}
    \footnotesize
    \begin{webassembly}[numbers=left, frame=l, framesep=1em]
(module
  (func $test
    (param $y i32) 
    (param $z i32) 
    (result i32)
    call $source
    if (result i32)
        local.get $y
        i32.const 2
        i32.add
    else
        local.get $z
        i32.const 3
        i32.mul
    end
    i32.const 1
    i32.add))
    \end{webassembly}
    
    \end{subfigure}
        \centering
    \begin{subfigure}[t]{0.67\columnwidth}
        \includegraphics[width=\textwidth]{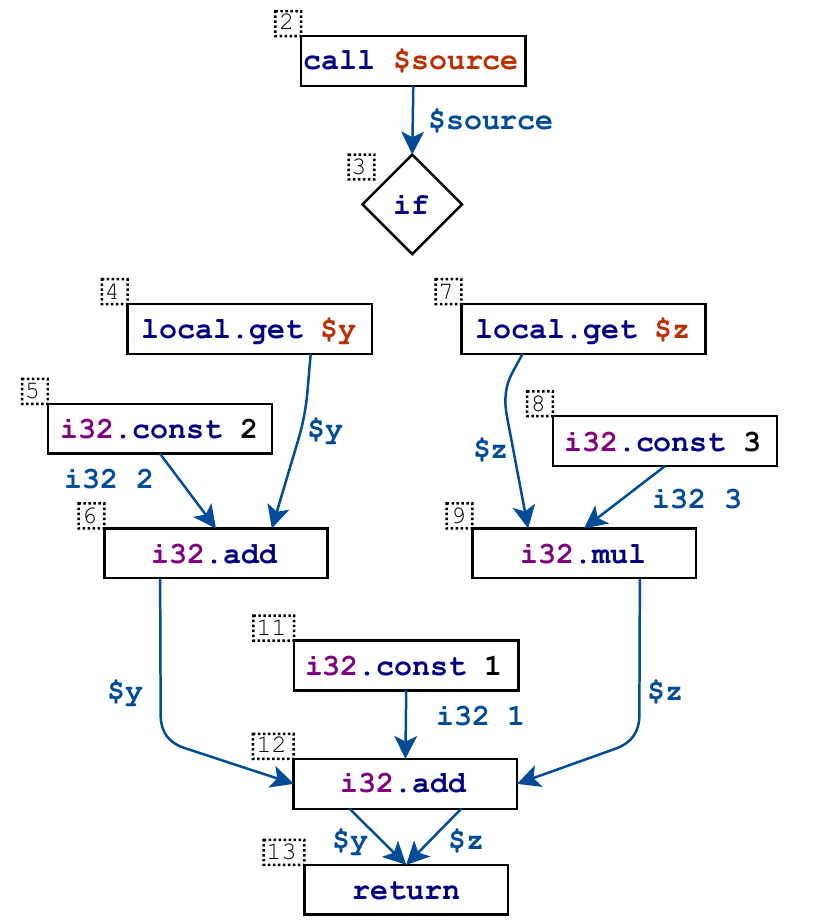}
    \end{subfigure}
    \caption{\small Simple \wasm program and its DDG}
    \label{fig:dep:grap:ex}
    \vspace{-0.3cm}
\end{figure}

Hence, for building \wasm data dependencies we employ two specific adaptations: i) we discard the calculation of control dependencies as it is easily queried later through the CFG, and ii) use a custom-made dependency analysis that combines reaching-definitions and constant/function propagation. This leads to our definition of Data Dependency Graph (DDG). More concretely, we track four types of data-dependencies: 
(1) a \emph{constant dependency} $\constdep{\id}{v, t}$ represents a data-dependency on a constant value $v$ of type $t$, generated by the instruction with identifier $\id$; 
(2) a \emph{function dependency} $\funcdep{\id}{f}$ represents a data-dependency on the value returned by function $f$, called at instruction with identifier $\id$;
(3) a \emph{global dependency} $\globdep{\id}{x}$ represents a data-dependency on the value of a global variable $x$ (through the instruction with identifier $\id$); and
(4) a \emph{local dependency} $\localdep{\id}{x}$ represents a data-dependency on the value of a local variable $x$. 
In the following, we use $\dep$ to range over the set of dependencies ($\dep \in \deps$) and $\depset$ to denote an arbitrary set of dependencies ($\depset \subseteq \deps$). Put formally: 
\begin{equation}
\begin{array}{lll}
\dep \in \deps & := & \constdep{\id}{v, t} \mid 
 \funcdep{\id}{f} \mid 
 \globdep{\id}{x} \mid 
 \localdep{\id}{x}
\end{array} 
\end{equation}
where: $v$ ranges over the set of \wasm values, $t$ over the set of \wasm types, $f$ over the set of function identifiers, and $x$ over the set of global and local variable~names.  

To better understand how dependencies are modeled, let us consider the \wasm program and DDG given in Figure~\ref{fig:dep:grap:ex}. Each node's identifier is displayed in its upper-left corner.
In particular, one can see that instruction $6$ depends on the value of the local variable \verb|$y|, which is put on top of the stack by instruction $4$. This dependency is modelled as: $\localdep{4}{\mathtt{\$y}}$. 
One can further see that instruction $6$ also depends on the constant value $2$ put on top of the stack by instruction $5$. This dependency is modelled as: $\constdep{5}{2, \sint}$. 

Dependencies are computed by a data-flow analysis explained next. Having computed the dependencies of every instruction, the construction of the DDG is straightforward. Each node is simply connected to the nodes on which it depends and DDG edges are labelled with attributes of their corresponding dependencies. For instance, 
let $e_1$ be the edge connecting nodes $4$ and $6$ in the DDG given in Figure~\ref{fig:dep:grap:ex}, we have that:
$\mu(e_1, type) = \textsc{DDG}$, 
$\mu(e_1, \text{\emph{ddgType}}) = \textsc{Local}$, and 
$\mu(e_1, \text{\emph{vName}}) = \mathtt{\$y}$. 
A complete description of the property-value pairs of DDGs is given in the appendix.

\subsection{Dataflow Analysis for \wasm}
\label{subsec:dataflow}

The construction of the DDG is the most challenging of all the CPG's sub-graphs as it requires a data flow analysis that takes into account the specific features of \wasm. In this section, we formally explain the steps and calculations implemented to track the data dependencies necessary for the DDG use case. As discussed above, despite applying known dataflow analysis, these data dependencies had to be adapted to our model and \wasm itself.

We apply the monotone framework to compute \wasm dependencies by lifting concrete states to abstract states. While a concrete state is composed of a concrete linear memory,  global store, local store, and stack. Our abstract states do not have the linear memory component as we do not track dependencies that are established through the use of linear memory operations. Instead, abstract states are simply composed of an abstract global store, local store, and stack.
In a nutshell, an abstract global store $\gsto : \globvars \rightharpoonup \wp(\deps)$ is a mapping from global variables to sets of dependencies (\emph{mutatis mutandis} for local stores: $\lsto : \lvars \rightharpoonup \wp(\deps)$). Analogously, an abstract stack, $\stack$, is simply a list of sets of abstract dependencies. For instance, $\gsto(\mathtt{\$g0}) = \constdep{20}{3, \sint}$ means that $\mathtt{\$g0}$ depends on the constant value $3$ via the instruction with identifier $20$.

To calculate the data dependencies at each execution point, we traverse the CFG of the program to be analyzed, propagating dependencies in a forward manner.
More concretely, we define, for each instruction, a transfer function that describes how that instruction propagates data dependencies by specifying how the output data dependencies are computed using the input data dependencies. 
Put formally, we define
a general transfer function $\transfer : \insts \times \astates \rightharpoonup \astates$ that computes an output  abstract state given an instruction in $I$ and an input abstract state. 
The function $\mathcal{T}$ is defined as a set of rules that follow the syntax of instructions. Below, we show a few selected rules, while the complete set of rules is given in the~appendix. Transfer function rules are annotated with the identifier of the executing instruction, \verb|id|. Their interpretation is straightforward. For instance, the transfer function for a \verb|const| instruction extends the stack with a constant value dependency.  

\begin{display}{Transfer functions (fragment): $\transfer : \insts \times \astates \rightharpoonup \astates$}
$
\begin{array}{l}
\transrule{t.const \ c}{(\gsto,\lsto,\stack)}{(\gsto,\lsto,\stack::\constdep{\id}{c, t})}{\id}
\\[3pt]
\transrule{local.get \ x}{(\gsto,\lsto,\stack)}{(\gsto,\lsto,\stack::\lsto(\tvar{x}))}{\id}
\\[3pt]
\transrule{local.set \ x}{(\gsto,\lsto,\depset::\stack)}{(\gsto,\lsto[\tvar{x} \mapsto \depset],\stack)}{\id}
\\[3pt]
\transrule{global.get \ x}{(\gsto,\lsto,\stack)}{(\gsto,\lsto,\stack::\gsto(\tvar{x}))}{\id}
\\[3pt]
\transrule{global.set \ x}{(\gsto,\lsto,\depset::\stack)}{(\gsto[\tvar{x} \mapsto \depset],\lsto,\stack)}{\id}
\end{array} 
$
\end{display}

Given the transfer functions, we compute the dependencies of each instruction by traversing the CFG of the program starting from the entry point of each function. Loop instructions are re-visited if their input dependencies change. The complete algorithm and the full list of transfer functions are given in the appendix. 

\subsection{Algorithmic Complexity}

We analyze the complexity of our algorithm for 
constructing WebAssembly CPGs by focusing on each sub-graph at a time.

The construction of the AST graph is done in linear time in the number of instructions of the original program. The complexity of the construction algorithm is dominated by the cost of the AST folding algorithm given in the appendix (Algorithm~\ref{algorithm:ast_folding}). Using aggregate analysis, we conclude that the total number of iterations of both the outermost and the innermost loops is linear in the number of instructions. In particular, we note that each node has at most one parent in the AST graph, meaning that it can only be visited once by the innermost loop.  

The construction of both the CFG and the CG graphs is done in linear time in the  number of instructions of the original program. The construction algorithms simply  traverse the instruction nodes of the program; the CFG algorithm connects each node to its immediate successors, while the CG algorithm connects direct and indirect call nodes to their corresponding functions. In both cases, the processing of each instruction is done in constant~time, making both algorithms linear in the number of traversed instructions. 

The construction of the DDG is done in quadratic time in the number of instructions of the original program. The cost of this algorithm is dominated by the dataflow analysis described in Section~\ref{subsec:dataflow}. 
To determine the cost of the dataflow analysis, we have to determine an upper-bound on height of the state lattice, which bounds the number of iterations that the dataflow analysis can perform~\cite{moller:book:2020}.
Recalling that each state is composed of an abstract global store $\gsto : \globvars \rightharpoonup \wp(\deps)$, an abstract local store $\lsto : \lvars \rightharpoonup \wp(\deps)$, and an abstract stack $\stack$, we note that the height of the state lattice corresponds  to the joint size of the domains of the three state components ($\vert \globvars \vert + \vert \lvars \vert + ST_{size}$) times the height of  the dependency lattice ($\wp(\deps)$); put formally: 
\begin{equation}
H = (\vert \globvars \vert + \vert \lvars \vert + ST_{size}) \times \vert \deps \vert    
\end{equation}
where $ST_{size}$ denotes an upper bound on the size of the stack. Observing that both $\vert \globvars \vert + \vert \lvars \vert + ST_{size}$ and $\vert \deps \vert$ are bounded by the total number of instructions, we conclude that the construction of the DDG is done in quadratic time.

\subsection{Optimizations}

The distinct structures that make up a CPG amount to a high number of nodes and edges. This is especially true for \wasm, which is a low-level language with many instructions which tends to require the generation and analysis of a very large graph even for relatively small binaries. The high number of nodes and edges hampers the scalability of our system with respect to memory usage and computing time for both CPG generation and graph traversals (queries). For example, regarding CPG generation, the biggest bottleneck is the construction of the DDG, mainly in the analysis of loops, because it has to traverse the graph multiple times and calculate data dependencies until no changes are made to the set of dependencies. This is very expensive as we realized that typical \wasm programs contain multiple chained loops.

To improve the efficiency and scalability of Wasmati, we employed several optimizations to construct the CPG in useful time without overwhelming the memory usage. The most relevant were as follows:
1) We pre-compute, for each type signature in the program, the set of functions with that signature that can be called indirectly, thus simplifying the generation of the call graph;
2) We propagate dependencies in a modular fashion, meaning that we only compute the dependencies of the successor of a loop, after computing the dependencies of the loop itself; this requires us to keep track of loop exits during data-flow analysis;
3) We cache the dependencies of inner loops to avoid re-analyzing them during new iterations of the outer loops;
4) We avoid the implementation of recursion as it had a negative impact due to indirection and high number call stack frames (eventually running out of stack memory). Every traversal/calculation was made in an iterative version.

\section{Querying WebAssembly CPGs}
\label{sec:queries}

\newcommand{\wasmati}[0]{Wasmati\xspace}

Vulnerabilities often give rise to specific CPG patterns that can be found using graph queries. Hence, the detection of a security flaw in \wasm can be achieved by specifying a graph query that captures the corresponding pattern. \wasmati comes with four query engine pipelines that can process queries in specified different languages. Next, we present WQL: Wasmati's dedicated query language for specifying and executing CPG queries. Then we give a brief account of the remaining query back-ends, explaining the rationale for their development and describing how they work. 
 
\subsection{\wasmati Query Language (WQL)} 

We introduce WQL by example, explaining how it can be used to encode and detect typical C/C++-style security vulnerabilities, namely \emph{use-after-free} vulnerabilities, \emph{taint-style} vulnerabilities, and \emph{buffer overflows}. Other common security vulnerabilities, such as use of \textit{dangerous functions}, \textit{format strings}, and \textit{double free} can also be easily expressed using WQL. The complete list of queries used to assess \wasmati is given in the appendix. 

\mypara{WQL features:} WQL is a simple interpreted imperative language that offers a wide range of built-in functions for traversing and inspecting the underlying CPG. WQL supports the standard primitive types: booleans, floats, integers, and strings. It also supports polymorphic lists and maps, as well as CPG nodes and edges. When it comes to control flow, WQL includes the typical control flow constructs: \texttt{if-then-else}, \texttt{while}, \texttt{foreach}, \texttt{break}, and \texttt{continue}. Furthermore, WQL has a dedicated syntax for range-expression, which is often useful for concisely implementing queries. We write \verb|[ n in lst : pred ]| to denote the list obtained by removing from \verb|lst| all its elements that do not satisfy \verb|pred|. 

\mypara{Use-after-free vulnerabilities} occur when one uses a reference to a memory location that has already been freed. This may lead to undefined system behavior and, in many cases, to a write-what-where condition. It can compromise the integrity and/or availability of the system by causing it to crash or lead to the corruption of valid data when that memory area has already been re-allocated.

Figure~\ref{fig:uaf} shows the WQL query for detecting use-after-frees. 
Our goal is to find three nodes \texttt{n1}, \texttt{n2}, and \texttt{n3} such that: 
(1) \texttt{n1} holds a call to \texttt{malloc}; 
(2) \texttt{n2} holds a call to \texttt{free}, which frees the memory segment allocated at \texttt{n1}; and 
(3) \texttt{n3} holds any instruction that executes after \texttt{n2} and uses the pointer returned by the instruction at \texttt{n1}.
In order to find such three nodes, we iterate over each function in the CPG. For each function, we first obtain all possible candidates for \texttt{n1} (line 2). For each possible \texttt{n1}, we then obtain all possible candidates for \texttt{n2} (line 4). In order to do this, we make use of the predicate \texttt{reachesDDG(n1, n, ``Function'', ``\$malloc'')} to identify only the nodes that depend on \texttt{n1} via the value returned by \texttt{\$malloc}. Finally, we use the built-in function \texttt{descendantsCFG} to obtain all the descendants of \texttt{n2} and then filter these descendants to obtain only those that depend on \texttt{n1} (line 6); if the resulting list is not empty, a \emph{use-after-free} vulnerability is flagged. 

\mypara{Taint-style vulnerabilities} refer to data flows from attacker-con-trolled sources to security-sensitive sinks that do not undergo sanitization.
With our CPGs, we can identify such flows simply by checking the data dependencies of all the possible sinks. 
Figure~\ref{fig:taint_func_to_func} shows the WQL query for detecting taint-style vulnerabilities. Our goal is to find the sink nodes that depend on source nodes. To this end, we inspect the dependencies of each sink, checking if any of them depends on a sensitive source. If the obtained list is not empty, a vulnerability is flagged for each individual source-sink pair.

\begin{figure}
    \centering
    \begin{wql}[numbers=left, frame=l]
foreach func in functions():
    nodes := [n in instructions(func) : (n.instType = "Call") && (n.label = "$malloc")];
    foreach n1 in nodes:
        descendants := [n in descendantsCFG(n1) : (n.instType = "Call") && (n.label = "$free") &&  reachesDDG(n1, n, "Function", "$malloc")];
        foreach n2 in descendants:
            uafs := [n in descendantsCFG(n2) : reachesDDG(n1, n, "Function", "$malloc")];
            if (!uafs.empty()) 
                vulnerability("Use after free", func.name, "$free");
    \end{wql}
    
    \caption{\small Use-after-free WQL Query.}
    \label{fig:uaf}
    \vspace{-0.3cm}
\end{figure}

\begin{figure}
    \centering
    \begin{wql}[numbers=left, frame=l]
foreach func in functions():
    sinkCalls := [n in instructions(func) : n.instType = "Call" && n.label in sinks &&   
                    !([e in n.inEdges : e.type = "DDG" && e.ddgType = "Function" &&                                                             e.label in sources].empty())]; 
    foreach sink in sinkCalls:
        vulnerability("Tainted", func.name, sink.label);
    \end{wql}
    
    \caption{\small Taint-flow WQL Query.}
    \label{fig:taint_func_to_func}
    \vspace{-0.3cm}
\end{figure}

\mypara{Buffer overflows:}
Despite having long been identified as a potential source of security vulnerabilities, buffer overflows remain a common gateway for attackers in today's code. 
Many mitigation techniques are in place to prevent buffer overruns, such as the use of stack guards to protect local data.
Unfortunately, no such technique has been implemented by the existing \wasm compilers.  

While buffer overflows can happen in various ways, many actually happen when handling strings inside loops. 
Figure~\ref{code:libpng} exemplifies one occurring inside a loop in a PNM decoding procedure in the module \verb|pnm2png| from the \verb|libpng| library (CVE-2018-14550)~\cite{CVE-2018-14550}.
The loop reads characters from a file and stores them in the buffer \verb|token| until the character read is one of those given in line 7. 
The problem is that the loop can have an arbitrary number of iterations, while the buffer has a fixed size. 
As a result, we can trigger a buffer overflow by picking a file containing a large enough string. 

The WQL query used to find this vulnerability is given in the appendix.
In a nutshell, 
we have to search for loop instructions for which: 
(1) there is a local variable \verb|$i| representing an index being incremented inside the loop;
(2) there is a store instruction within the body of the loop that depends on \verb|$i| (buffer write operation); and 
(3) there is no explicit loop exit (\verb|br_if|) instruction that depends on the result of a comparison operation directly involving the value of \verb|$i| (e.g. \verb|i < BUF_SIZE|).
For (1), we look for \verb|i32.add| instructions with two incoming DDG edges: one expressing a \emph{local dependency} on variable \verb|$i| and another expressing a constant dependency on the value to be added. 
For (2), we simply look for \verb|store| instructions that directly use the result of the \verb|i32.add| instructions found in (1). 
Finally, for (3), we search for \verb|br_if| instructions that rely on the result of a comparison operation involving the value of \verb|$i|. 
If no such instructions are found, a vulnerability is~flagged. 

\begin{table*}[t]
\centering
{\footnotesize
\resizebox{0.9\textwidth}{!}{%
\begin{tabular}{clcccc|ccc|ccc|ccc|ccc} 
\toprule
\multirow{2}{*}{ \textbf{\#} } & \multicolumn{1}{c}{\multirow{2}{*}{\textbf{Query Description} }} & \multicolumn{4}{c|}{ \textbf{LOC} } & \multicolumn{3}{c|}{\textbf{Basic} } & \multicolumn{3}{c|}{\textbf{Lehmann} } & \multicolumn{3}{c|}{\textbf{STT} } & \multicolumn{3}{c}{\textbf{CWE} } \\ 
\cmidrule{3-18}
 & \multicolumn{1}{c}{} & \textbf{NAT} & \textbf{WQL} & \textbf{N4J} & \multicolumn{1}{c|}{\textbf{DTL}} & \multicolumn{1}{c}{\textbf{TP} } & \multicolumn{1}{c}{\textbf{FP} } & \multicolumn{1}{c|}{\textbf{P} } & \multicolumn{1}{c}{\textbf{TP} } & \multicolumn{1}{c}{\textbf{FP} } & \multicolumn{1}{c|}{\textbf{P} } & \multicolumn{1}{c}{\textbf{TP} } & \multicolumn{1}{c}{\textbf{FP} } & \multicolumn{1}{c|}{\textbf{P} } & \multicolumn{1}{c}{\textbf{TP} } & \multicolumn{1}{c}{\textbf{FP} } & \multicolumn{1}{c}{\textbf{P} } \\
\rowcolor[rgb]{0.851,0.851,0.851} 1 & Format Strings & 24 & 10 & 19 & 139 & 5 & 0 & 7 & - & - & - & 11 & 8 & 11 & 2 & 0 & 2 \\
2 & Dangerous Function & 14 & 8 & 5 & 137 & 7 & 0 & 7 & 1 & 0 & 1 & 10 & 0 & 10 & 2 & 0 & 2 \\
\rowcolor[rgb]{0.851,0.851,0.851} 3 & Use After Free & 67 & 15 & 16 & 140 & 2 & 0 & 2 & - & - & - & - & - & - & - & - & - \\
4 & Double Free & 60 & 16 & 19 & 141 & 1 & 0 & 1 & - & - & - & - & - & - & - & - & - \\
\rowcolor[rgb]{0.851,0.851,0.851} 5 & Tainted CallIndirect & 55 & 12 & 13 & 139 & - & - & - & - & - & - & 5 & 0 & 5 & - & - & - \\
6 & Tainted Func-to-Func & 55 & 12 & 13 & 142 & 2 & 0 & 2 & - & - & - & - & - & - & 1 & 0 & 1 \\
\rowcolor[rgb]{0.851,0.851,0.851} 7 & Tainted Local-to-Func & 84 & 51 & 55 & 177 & 16 & 0 & 16 & 4 & 0 & 4 & - & - & - & 4 & 0 & 4 \\
8 & BO - Static Buffer & 141 & 58 & 58 & 172 & 1 & 0 & 2 & 0 & 0 & 1 & 12 & 0 & 15 & 8 & 0 & 9 \\
\rowcolor[rgb]{0.851,0.851,0.851} 9 & BO - Static Buffer (malloc) & 86 & 27 & 19 & 142 & 1 & 0 & 1 & - & - & - & - & - & - & 1 & 0 & 1 \\
10 & BO - Loops & 94 & 23 & 47 & 147 & - & - & - & 1 & 0 & 1 & 2 & 0 & 2 & 1 & 0 & 1 \\ 
\cmidrule{1-18}
\multicolumn{1}{r}{} & \multicolumn{1}{r}{\textbf{Total:} } & \textbf{680} & \textbf{232} & \textbf{264} & \textbf{1476} & \textbf{35}  & \textbf{0}  & \textbf{38}  & \textbf{6}  & \textbf{0}  & \textbf{7}  & \textbf{40}  & \textbf{8}  & \textbf{43}  & \textbf{19}  & \textbf{0}  & \textbf{20}  \\
\bottomrule
\end{tabular}}}
\caption{\small Query vulnerability report by dataset and LOC metrics for each querying approach. Acronyms: BO (Buffer Overflow), NAT (native, C++), WQL (Wasmati Query Language), N4J (Neo4J Cypher) and DTL (Datalog).}
\label{table:query_report}%
\vspace{-0.6cm}
\end{table*}

\subsection{Other Query Back-ends} 

By developing multiple query engine back-ends, our primary goal was to investigate the trade-offs that different query language paradigms can offer regarding: i) expressiveness power, ii) conciseness and simplicity, and iii) query execution performance. After developing and experimentally analyzing three query back-ends that support imperative, logic, and database processing paradigms -- C++, Datalog, and Neo4j, respectively -- we developed our own domain-specific language (WSL) which i) offers enough expressiveness for capturing \wasm vulnerability patterns, ii) is relatively intuitive to program, and iii) performs well. Next, we review the remaining three query back-ends, explaining their main properties and how they can be used to query \wasm CPGs.

\mypara{Native back-end:}
\wasmati comes with an API for users to implement their queries directly in C++ and integrate them within the \wasmati code base. To this end, a user simply needs to write a file containing a C++ function that describes their query and then adds the query to a configuration file with all the queries to be executed.
A complete description of our query API is given in the appendix. 
By writing queries natively, the user can leverage the high performance and expressiveness of C++.
However, users have to re-compile \wasmati every time they need to change or add a query to the system and they have to be acquainted with both C++ and the structure of the \wasmati code-base. The implementation of some search patterns for Wasm code vulnerabilities (e.g., buffer overflows) is also rather complex and prone to programming errors.

\begin{figure} [t]
	\centering
	\footnotesize
	\begin{neo4j}[numbers=left, frame=l]
MATCH (f:Function)-[:AST*1..]->(sink:Instruction)
WHERE sink.instType="Call" AND sink.label IN sinks

WITH * MATCH (src:Instruction)-[:DDG*1..]->(sink)
WHERE src.instType="Call" AND src.label IN sources
    AND (source_call)-[:DDG*1.. {ddgType:"Function", label:src.label}]->(sink)
RETURN src.label as source, sink.label as sink, 
    f.name as function;
\end{neo4j}
\vspace{-0.3cm}
\caption{\small Tainted source-to-sink in Neo4j CQL.}
\label{code:neo4j}
\vspace{-0.3cm}
\end{figure}

\begin{figure} [t]
	\centering
	\footnotesize
	\begin{datalog}[numbers=left, frame=l]
taintedFuncToFunc(FUNC_NAME, Y, SINK) :-
    sources(SOURCE), call(X, SOURCE, _, _), 
    reachesFunc(FUNC_NAME, X),
    sinks(SINK), call(Y, SINK, _, _), 
    reachesDDG(X, Y, "Function", SOURCE).
\end{datalog}
\vspace{-0.3cm}
\caption{\small Tainted source-to-sink in Datalog.}
\label{code:datalog}
\vspace{-0.3cm}
\end{figure}

\mypara{Neo4j:}
The CPG can be automatically loaded into a Neo4j~\cite{neo4j} database and running graph queries specified in Cypher, an SQL-like language.  Figure~\ref{code:neo4j} shows the Cypher query for finding taint-style vulnerabilities. The execution of lines 1 and 2 yields an interim table of pairs, each consisting of a function node and one of its sink calls. 
In lines 4-6, we look for source calls that reach any of the sinks stored in the interim table via DDG function edges.  
Finally, in line 7, we return a table with the computed results, with each result corresponding to a taint-style vulnerability (source name, sink name, enclosing function name). 
Neo4j relies on sophisticated planning algorithms for maximizing the query execution performance.
From our experience, however, query planning works better for simpler queries, as we have observed significant performance degradation for complex queries involving multiple graph traversals.

\mypara{Datalog:}
The CPG can also be automatically loaded into Souffé~\cite{jordan2016souffle}, a state-of-the-art Datalog engine.
Wasmati also provides a library containing various predicates to reason about the structure of \wasm CPGs in a declarative fashion. For instance, it comes with a predicate  \verb|reachesDDG(X, Y, TYPE_DEP, LAB)| to denote that the node \verb|Y| is reachable from the node \verb|X| via DDG edges with label \verb|LAB| of type \verb|TYPE_DEP|. 
Datalog queries are expressed as predicates for which the Datalog solver will try to find a model.  
When defining new query predicates, the user can leverage our library of CPG predicates which can result in a rather concise query specification. Figure~\ref{code:datalog} shows the Datalog query for finding taint-style vulnerabilities:  
we find a call to a source and a call to a sink using the predicates \verb|sources| and \verb|call| and then we require that the sink be reachable from the source using DDG function edges. On the other hand, Datalog is generally the less performing back-end.

\section{Evaluation}
\label{sec:eval}

This section presents our experiments to evaluate Wasmati. They mainly focus on answering the following questions:

\begin{enumerate}
    \itemsep0em 
    \item Can security vulnerabilities in WebAssembly code be modeled and located using a CPG?
    \item Does Wasmati find security vulnerabilities in WebAssembly code collected in the wild?
    \item How well does Wasmati scale when generating CPGs from large real-world applications?
    \item How do the different graph querying back-ends of Wasmati scale over CPGs generated from large applications?
\end{enumerate}

Next, we present our experimental setup and main findings for each of these questions in a separate section.

\subsection{Evaluation Using Annotated Datasets}

In the first part of our evaluation of Wasmati, our goal is to assess the feasibility of modeling and finding security vulnerabilities in \wasm using CPGs. To achieve this, we implemented and executed the ten different queries described in $\S$\ref{sec:queries} over four datasets containing programs written in C with known vulnerabilities.
We compiled the datasets to WebAssembly using Emscripten 2.0.9~\cite{emscripten} with level 1 optimization and debug information. 
For our dataset selection, the vulnerabilities therein contained had to be properly annotated, even if the number of programs in the dataset was relatively small, as this allows us to access the ground truth and evaluate the detection effectiveness of our system. Next, we describe the four datasets that we used, containing a total of 110 C programs.

\begin{enumerate}
    \itemsep0em 
    \item \textbf{Basic:} has a total of 37 C programs compiled and created by us during the implementation and testing of Wasmati.
    
    \item \textbf{Lehmann:} contains 7 programs from Lehman et al.~\cite{lehmann2020everything}. The programs are attack primitives and end-to-end exploits. It includes a program using the vulnerable version of \verb|pnm2png| from libpng depicted in Figure~\ref{code:libpng}, a remote execution code in NodeJS and arbitrary file write.
    
    \item \textbf{STT:} is a repository of 47 C vulnerable programs from Binary Analysis School~\cite{stt}.
    The programs are exploit exercises following the style of Capture the Flag.
    
    \item \textbf{CWE:} is a list of ``Weaknesses in Software Written in C'' with a total of 19 example code snippets~\cite{cwe_c}. We removed the weaknesses where no query existed targeting it or the vulnerability in C is not ported to WebAssembly. We also removed duplicated code. One removed example is CWE-467 which flags the use of \verb|sizeof()| on a pointer type.

\end{enumerate}

\mypara{Query expressiveness:} We found that all our queries can be encoded using the language primitives provided by C++ (native), WQL, and Datalog. As for Neo4j, queries 7 and 8 required us to develop specific workarounds. The simplicity of Neo4j's Cypher querying language shines in querying for simple logical patterns. However, since it is a declarative language, it does not natively support recursion, which is quite useful for query 7. Cypher also lacks helper data structures like lists or maps that are convenient for query 8. To overcome these limitations, we employed different approaches. For query 7, we emulated recursion by 
recurrently invoking a simpler query through the Neo4j Python driver. For query 8, we developed a simple aggregation plugin function in Java that builds a map of local buffer indexes and their corresponding buffer sizes. For these reasons we consider Cypher's expressiveness to be relatively more limited for representing our queries than the other query back-ends.

\mypara{Query conciseness:} Table~\ref{table:query_report} presents the lines of code (LOC) of each query written in the four query specification languages supported by Wasmati: C++, WQL, Neo4j, and Datalog. The Datalog queries include a shared library that contains common predicates. The size of this library is 127 LOC. If we exclude this library from the total size (i.e., 1476) and count only the lines of code specific to each query, then the total query size is 206 LOC. We can then see that WQL, Neo4j, and Datalog allow for writing queries with comparable levels of conciseness, averaging respectively 23.2, 26.4, and 20.6 LOC per query. As expected, C++ is considerably more verbose displaying a LOC size 3$\times$ larger than the other languages.

\mypara{Query effectiveness:} To gauge how effective our queries are at finding vulnerabilities in our four datasets, Table~\ref{table:query_report} reports on the absolute number of \textit{true positives} (TP), \textit{false positives} (FP) and the number of \textit{present} vulnerabilities in the code (P), i.e., the ground truth. We count a TP as a correctly reported vulnerability and a FP otherwise. 
We categorize each present vulnerability according to its nature and assign it to the query that best matches its pattern.

From a total of 108 vulnerabilities in 110 programs, 100 were correctly reported (TP) and 8 were false positives (FP). The false-positive rate, which computes the proportion of incorrect reports (FP) in relation to the total results (P), is 7.41\%. The precision, which reflects the proportion of true positives in the total reports made by the queries, is 92.59\%. There are also 8 undetected vulnerabilities (FN): 3 in Basic, 1 in Lehmann, 3 in STT, and 1 in CWE. These anomalies, both in missed vulnerabilities and ill-classified ones, occur only in format strings (query 1) and buffer overflows in static buffers (query 8). All other queries have detected all existing vulnerabilities in the datasets while flagging no false positives. Consequently, Wasmati's recall for these datasets is 92.59\%. This means that Wasmati achieves both high recall and high precision, which shows its effectiveness for detecting vulnerabilities in WebAssembly binaries.

In the case of query 8, 
two unreported vulnerabilities in the basic dataset 
arise from calls to \verb|printf()| using global static buffers as its first argument.
In WebAssembly, both constant data strings and global static buffers are indistinguishable from each other and, as result, the vulnerability is not reported as such. In the STT dataset, a total of 8 false positives were reported by the query. Analyzing the binary, we find that when there are multiple \verb|printf| calls sequentially. The way that the WebAssembly compiler stores the constant format templates leads the query to wrongly perceive its logic. 
As for the buffer overflows in static buffers, a total of 4 (all in global static buffers) were not reported. The buffer is stored in the data section and referred by a constant pointer. The query cannot infer the size of a global static buffer so, it 
omits the report.

\begin{table}[t]
\centering
\small
\begin{tabular}{l|cc|cc|cc|cc|cc} 
\toprule
& \multicolumn{2}{c|}{\textbf{Basic} } & \multicolumn{2}{c|}{\textbf{Lehmann} } & \multicolumn{2}{c|}{\textbf{STT} } & \multicolumn{2}{c}{\textbf{CWE} } & \multicolumn{2}{|c}{\textbf{Total} }\\ 
\cmidrule{2-11}
 \multicolumn{1}{c}{} & \multicolumn{1}{c}{\textbf{TP} } & \multicolumn{1}{c}{\textbf{FP} } & \multicolumn{1}{c}{\textbf{TP} } & \multicolumn{1}{c}{\textbf{FP} } & \multicolumn{1}{c}{\textbf{TP} } & \multicolumn{1}{c}{\textbf{FP} } & \multicolumn{1}{c}{\textbf{TP} } & \multicolumn{1}{c}{\textbf{FP} } & \multicolumn{1}{c}{\textbf{TP} } & \multicolumn{1}{c}{\textbf{FP} }\\
\rowcolor[rgb]{0.851,0.851,0.851} Wasmati & 35 & 0 & 6 & 0 & 40 & 8 & 19 & 0 & \textbf{100} & \textbf{8}\\
\rowcolor[rgb]{0.851,0.851,0.851} Joern & 10 & 17 & 0 & 6 & 24 & 38 & 4 & 4 & \textbf{38} & \textbf{65}\\
\bottomrule
\end{tabular}
\caption{\small Comparison of aggregate results between Wasmati (analysis of \wasm binary compiled from C code) and Joern (analysis of original C code).}
\label{table:query_report_joern}
\vspace{-0.6cm}
\end{table}

\mypara{Comparison with Joern:} The four datasets used to evaluate Wasmati contain programs written in C with known vulnerabilities. The results presented above show that these vulnerabilities persist in the corresponding \wasm binary after compiling the C code, and that Wasmati can successfully detect these vulnerabilities. To offer a comparison between Wasmati CPGs and the CPGs originally developed to detect vulnerabilities in C/C++ code~\cite{cpg}, we decided to perform an evaluation of Joern~\cite{joern}, which is the open-source implementation of the original published work that is actively maintained. Note that it is also the basis of a commercial tool developed by ShiftLeft~\cite{shiftleft}. Consequently, although Joern is actively maintained, we expect it not to include the latest detection features or perform on par with its commercial version. We used the queries shipping with Joern, which are aimed at detecting the most common vulnerabilities in C code, and executed Joern directly on the C files of each dataset. Table~\ref{table:query_report_joern} shows the aggregate results of Joern and Wasmati. From a total of 108 vulnerabilities in 110 programs, Wasmati correctly reported 100 vulnerabilities (TP) and reported 8 false positives (FP), while Joern correctly reported only 38 vulnerabilities (TP) and reported 65 false positives (FP). This further shows that CPGs can be successfully applied to \wasm for detecting vulnerabilities, with Wasmati achieving 92.59\% recall and 92.59\% precision, and Joern 35.19\% recall and 36.89\% precision when analyzing the original C files. We infer that this discrepancy in the results is directly linked to the coverage of the queries implemented by both tools and not necessarily mean that CPGs are less effective for C code. In Wasmati we focused on implementing comprehensive queries, while the open-source Joern implementation offers only 14 queries covering dangerous functions, format strings, buffer overflows, use-after-free, and other, corresponding to 101 out of 108 total vulnerabilities in the datasets.

\subsection{Vulnerability Detection in the Wild}

In this second part of our evaluation, our goal is to assess Wasmati using \wasm binaries deployed in the wild. To this end, under ideal conditions, we would like to test Wasmati on a dataset that: i) contains a representative and large collection of real-world \wasm binaries, and ii) comes with ground truth annotations that allow us to determine whether or not the vulnerabilities detected by Wasmati are real. In the absence of such an ideal dataset, we used the dataset curated by Hilbig et al.~\cite{hilbig2021empirical}. It consists of 8,461 unique binaries collected from several sources (repositories, package managers, and websites), and it is not annotated.

We started by generating the CPGs for all the \wasm binaries of this dataset. From the 8,461 binaries, Wasmati has successfully generated CPGs for 7,879 (93.1\%) binaries. Of the remaining 582 binaries, 561 (6.6\%) failed mainly due to unsupported \wasm features in the binaries (e.g. threading, multi-return values, bad sections, etc..) and 21 (0.3\%) exceeded the maximum allocated 16 GiB of RAM. These 21 programs had a mean size of 61 MiB. After gathering the generated CPG, we ran the 10 queries listed in Table~\ref{table:query_report}. Next, we present the main findings of our analysis.

\begin{table}
\footnotesize
\centering 
\begin{tabular}{lrr}
    \toprule
    \textbf{Deployment Location} & \textbf{Vuln. Binaries} & \textbf{Total Collected} \\ \hline 
    github & 3,761  (8.5\%) & 44,218 \\ 
    web/httparchive & 86  (33.0\%) & 261 \\ 
    web/own-crawler & 264  (9.0\%) & 2,923 \\ 
    npm/wasm & 254  (7.3\%) & 3,488 \\ 
    wapm & 23  (19.0\%) & 122 \\ 
    firefox-extensions & 7  (24.1\%) & 29 \\ 
    manual & 17  (35.4\%) & 48 \\ 
    survey & 9  (20.0\%) & 45 \\ 
    npm/top & 3  (21.4\%) & 14 \\
    \midrule
    \textbf{Total} & 4,424  (8.6\%) & 51,148 \\
    \textbf{Total (unique binaries)} & 3,140  (37.1\%) & 8,461 \\
    \bottomrule
\end{tabular}
\caption{\small Deployment location of binaries for which Wasmati detected at least one vulnerability. Percentage in relation to total number of collected binaries for that deployment location. Note that, in the dataset~\cite{hilbig2021empirical}, the same binary might be collected from different sources, which leads to duplicates across lines in the table.}
\label{tab:collection-methods}
\vspace{-0.8cm}
\end{table}

\mypara{Provenance of potentially vulnerable binaries:} Table~\ref{tab:collection-methods} shows the deployment location of \wasm binaries for which Wasmati detected at least one potential vulnerability (in the dataset~\cite{hilbig2021empirical} this information corresponds to the binary collection method). We see that the original dataset is heavily skewed, where 86.5\% of all binaries of the dataset originate from Github. This helps explain why most vulnerable binaries detected by Wasmati (3,761) were originally found in Github repositories. We also observed that many of these binaries were obtained from repositories owned by security researchers, who collect selected \wasm binaries for research purposes. Beyond Github, Wasmati detects vulnerabilities in a considerable percentage of binaries that reach the production stage, such as real websites (\textit{web/httparchive}), at the WebAssembly Package Manager (\textit{wapm}), in \textit{firefox-extensions} and at the Node Package Manager (\textit{npm/top}). This demonstrates the feasibility of applying Wasmati for the analysis of real \wasm binaries.

\mypara{Characterization of potentially vulnerable binaries}: Table~\ref{tab:vuln-metrics} shows the number of vulnerabilities detected by Wasmati in the curated dataset, characterized by vulnerability type. In total, Wasmati flags 83,202 potential vulnerabilities in 3,140 unique binaries deployed in the wild. We discuss three main observations:

\noindent \textit{1. Large number of taint-style vulnerabilities:} In our query for detecting this type of vulnerabilities,  
any data external to the \wasm binary is considered tainted. 
This query flags a potential vulnerability in two conditions: i) data dependent on a tainted input reaches a known dangerous sink inside the \wasm module (e.g. \texttt{memcpy}), or ii) data dependent on a tainted input is reflected to the calling environment (e.g. when the \wasm module calls a JavaScript function). Wasmati detects that such data flows can be frequent. A closer analysis revealed a heavily skewed distribution, with 73.6\% (59486) of the total tainted-variable vulnerabilities being located in 10\% of the flagged binaries. Furthermore, we found multiple binaries that are similar, as is the case of the first and second binaries with most tainted-variables flagged (3028 and 2973 respectively) that are compilations from libxml2. 
This library has many imported functions from JavaScript that handle HTML and XML, and are good candidates to consider as vulnerable sinks.

\noindent \textit{2. Usage of dangerous functions:} In Table~\ref{tab:vuln-metrics} we can see 156 uses of dangerous functions: \texttt{gets} (1 occurrence) and \texttt{strcat} (155 occurrences). These are generally considered dangerous regardless of how they are used~\cite{CWE-242}. In fact, some C/C++ compilers warn the developer when \texttt{gets} is used since it was removed from ISO/IEC 9899:2011~\cite{iso_iec_9899_2011}. Yet, these functions persist in \wasm code.

\begin{table}
\footnotesize
\centering
\begin{tabular}{lcccccc|c} 
    \toprule
    \textbf{Language} & \textbf{FS} & \textbf{TV} & \textbf{UAF} & \textbf{DFree} & \textbf{BO} & \textbf{DFunc} & \textbf{Binaries}  \\ \midrule
    C & 654 & 8941 & 6 & 5 & 90 & 64 & 49 \\ 
    Rust & - & 62 & - & - & 343 & - & 73 \\ 
    Unknown & 925 & 71821 & - & - & 199 & 92 & 3018 \\ 
\cmidrule{1-8}
    \textbf{Total:} & 1579 & 80824 & 6 & 5 & 632 & 156 & 3140 \\
\bottomrule
\end{tabular}
\caption{\small Number of vulnerabilities detected by Wasmati: format strings (FS), tainted variable (TV), use after free (UAF), double free (DFree), buffer overflow (BO) and dangerous function (DFunc).}
\label{tab:vuln-metrics}
\vspace{-0.6cm}
\end{table}

\mypara{Case studies}: Given that i) no tool still exists to automatically validate the vulnerabilities reported by Wasmati, and ii) the manual inspection of such a large \wasm codebase would be extremely laborious, we sampled a few vulnerabilities reported by Wasmati to analyze manually. From this exercise, we identified two interesting cases where reported buffer overflows may actually result in exploitable vulnerabilities unless the provided inputs are properly sanitized by external JavaScript code. In one case, we identified 4 binaries that use vulnerable versions of the libpng library. Wasmati detected the buffer overflow vulnerability in libpng reported in CVE-2018-14550~\cite{CVE-2018-14550} (see Figure~\ref{code:libpng}). This demonstrates the transfer of real vulnerabilities to \wasm in the wild. In a second case, we confirmed the existence of a buffer overflow in a \wasm binary that pertains to a Firefox Extension for a Cardano ADA Wallet.
Wasmati identifies that the function \verb|__wbg_publickey_free| is callable from JavaScript. Wasmati treats its parameter (a pointer) as tainted and tracks it as it reaches a sensitive allocation sink (\verb|memcpy|). This pointer does not undergo sanitization until it reaches \verb|memcpy|; if the value gets corrupted somehow in the JavaScript code, it can result in a possible buffer overwrite and memory corruption that can chain into other problems (e.g. code execution, or signature bypass/forgery).

\subsection{CPG Generation Performance}

In this section, we assess the scalability, overall performance, and resources allocated in the generation of the CPG.
To achieve these goals we generated CPGs from two well-known industrial benchmarks: PolybenchC~\cite{polybench} and SPEC CPU 2017~\cite{spec2017}. The experimental evaluation was performed in a 64bit Ubuntu 18.04LTS with 16GB RAM and 4 Intel Core i7-4700HQ 2.40GHz CPUs with all runs using the same configuration file, which specifies sources, sinks, etc.

\begin{table}[t]
\centering
\footnotesize
\begin{tabular}{lrr|r} 
\toprule
\textbf{Source} & \textbf{C} & \textbf{C++} & \textbf{Average (binary)} \\ 
\midrule
\textbf{Instruct. (k)} & 497.5k & 1M &  713.0k \\
\textbf{Size (KiB)} & 1,167 & 2,697 &  1,797 \\
\textbf{Nodes} & 525.7k & 1.1M &  747.6k \\
\textbf{Edges} & 1.6M & 9.7M &  4.9M \\
\textbf{Memory (MiB)} & 107.88 & 419.50 & 236.17 \\
\textbf{Time} & 37s & 1min 43.5s &  57.79s \\
\textbf{Exported (MiB)} & 8.74 & 42.01 & 22.44 \\
\bottomrule
\end{tabular}
\caption{\small Average CPG generation times and information from SPEC CPU 2017 binaries. Complete data by binary in appendix.} 
\label{tab:spec2017-summary}
\vspace{-0.7cm}
\end{table}

\mypara{Scalability assessment:} We generated CPGs for a subset of the SPEC CPU 2017 benchmark comprising a total of 17 binaries, from C and C++ source, compiled using Emscripten 2.09. The results are shown in Table~\ref{tab:spec2017-summary}. On average, the binaries are composed of 713k instructions with 502.gcc and 526.blender peeking in 2.9M and 3.2M instructions respectively.
The constructions of the CPG took an average of 58 seconds per binary with a total of about 16min22s. We can see that even the largest binaries generated their CPG in less than 4min30s using up to 1.73 GB of RAM.

\begin{figure}[t]
    \centering
    \includegraphics[width=0.8\columnwidth]{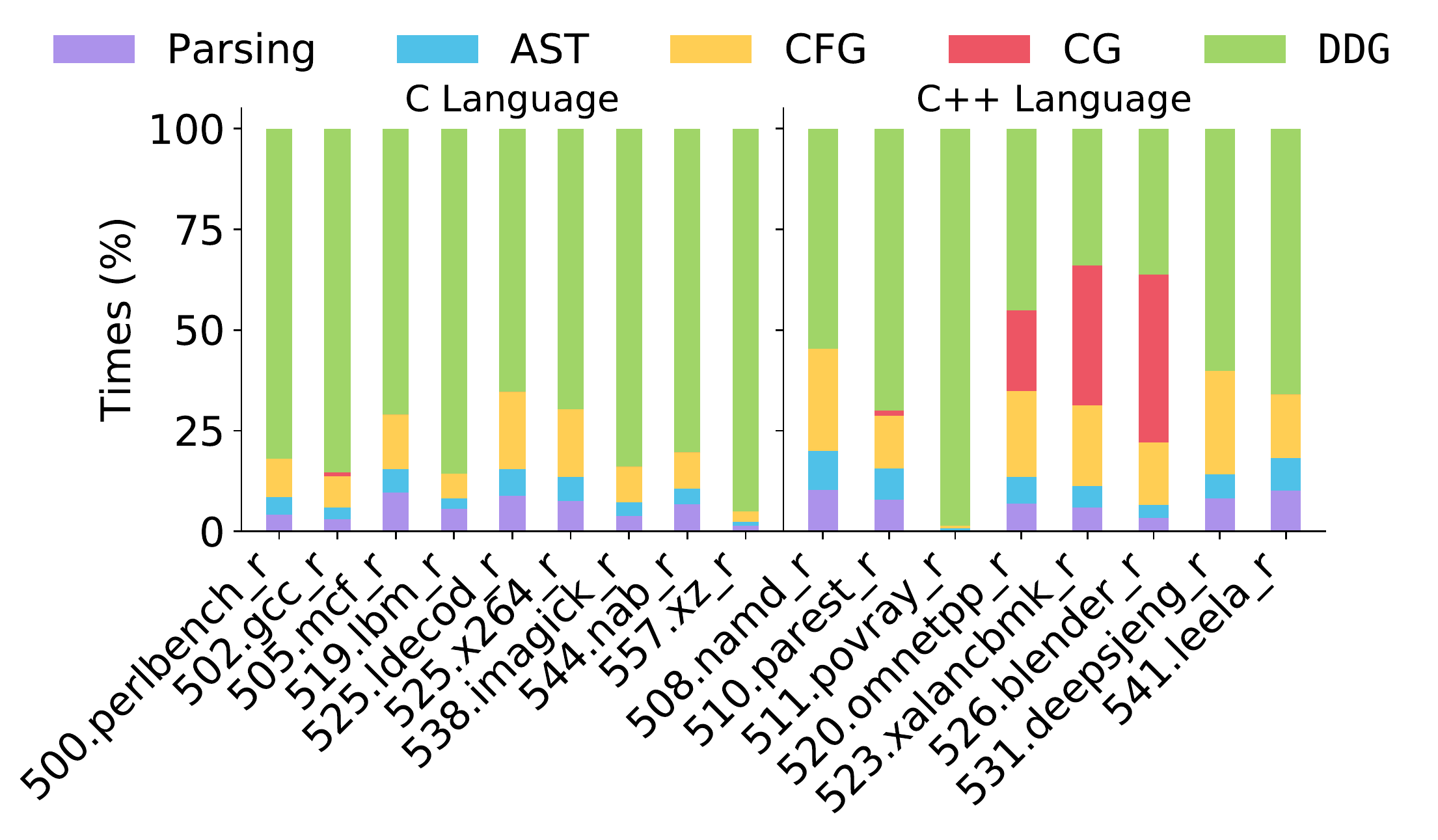}
    \vspace{-0.3cm}
    \caption{\small Time (\%) per CPG's construction stage.}
    \label{fig:time_bars}
    \vspace{-0.3cm}
\end{figure}

Figure~\ref{fig:time_bars} presents a more detailed analysis of the time taken to construct the CPG per benchmark, showing how much time is employed by each stage of the construction of the CPG, namely: parsing, construction of the AST, construction of the CFG, construction of the CG, and construction of the DDG.  As expected, the DDG construction captures the most time of the construction time. Some exceptions can be seen in 2 C++ programs for which most of the time of the CPG construction is spent on building the call graph (most likely due to the presence of many call indirect instructions).

\begin{figure}[t]
    \centering
    \includegraphics[width=0.6\columnwidth]{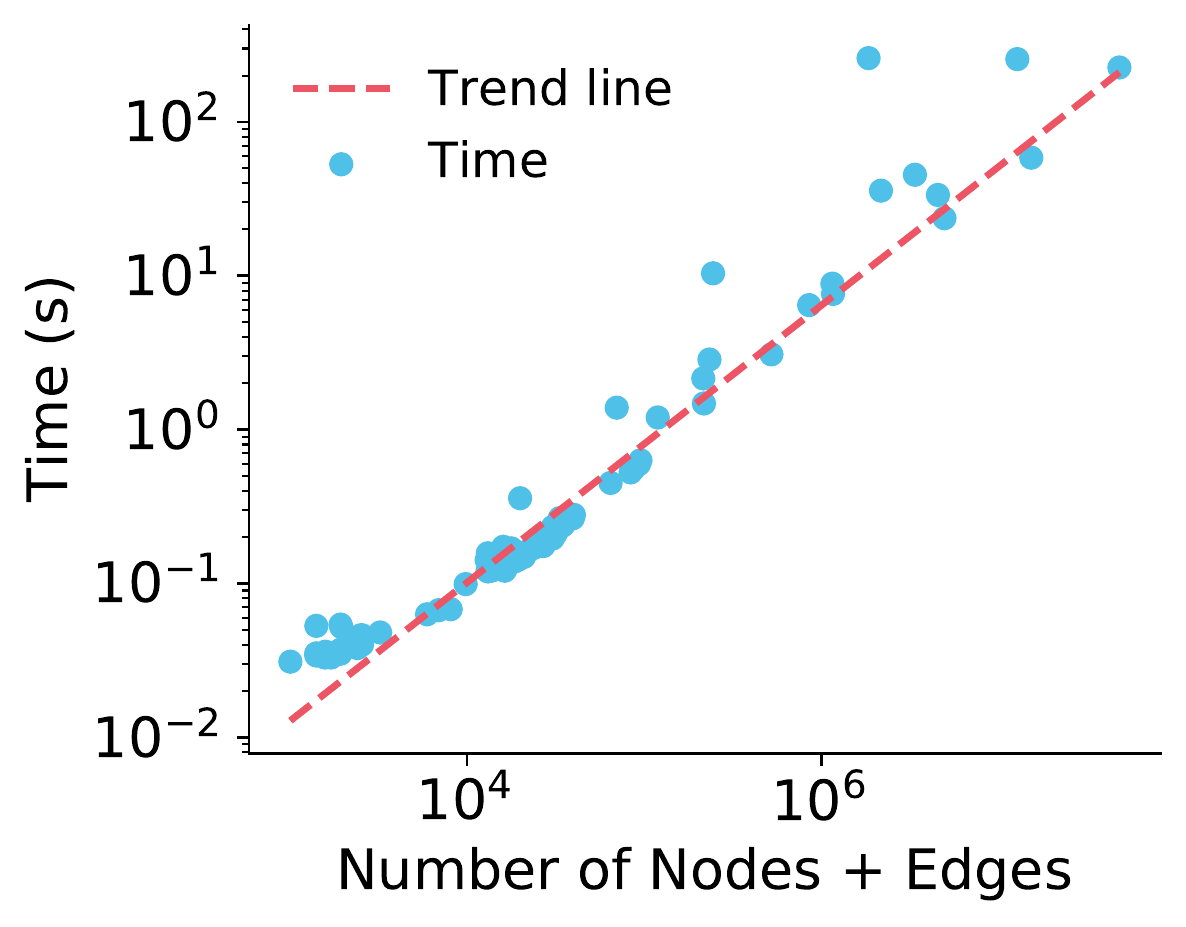}
    \vspace{-0.3cm}
    \caption{\small CPG construction time by size.}
    \label{fig:construction-cpg}
    \vspace{-0.4cm}
\end{figure}

Figure~\ref{fig:construction-cpg} represents the CPG construction time in terms of the graph size, counting the total number of nodes and edges. The plot includes the sizes of all 158 binaries from our four datasets and two benchmarks. The regression follows a power series with $R^2 = 0.929$. This result shows that the CPG construction is polynomial in function of the graph's size and can scale to larger programs.

\begin{figure}[t]
    \centering
    \hspace*{-0.9cm}\includegraphics[width=1.1\columnwidth]{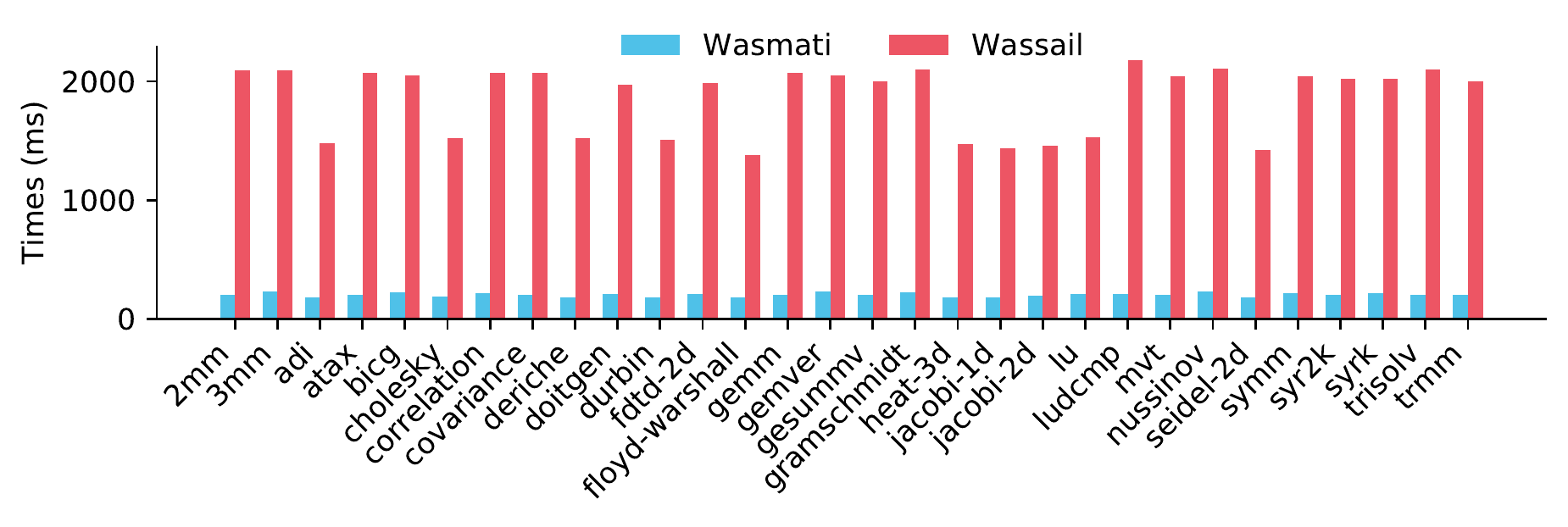}
    \vspace{-0.8cm}
    \caption{\small Wasmati's generation time comparing to Wassail.}
    \label{fig:wasmati_vs_wassail}
    \vspace{-0.1cm}
\end{figure}

\mypara{Comparison against related systems:} To put Wasmati's results in perspective, we compared it against the closest-related system in the literature named Wassail~\cite{stievenart2020compositional}. Wassail is a static taint analysis tool for WebAssembly programs focused exclusively on information flow analysis. In contrast to Wasmati, which generates full-blown CPGs, Wassail generates a much simpler graph data structure consisting only of the control flow graph (CFG) of the program followed by a data flow analysis propagation over the CFG. Wassail was written in OCaml and evaluated using the PolybenchC suite, which is composed of 30 C programs. To compare the performance between Wasmati and Wassail, we tested both systems against PolybenchC. The data structure computed by Wassail is comparable to the DDG generated by Wasmati. Figure~\ref{fig:wasmati_vs_wassail} displays the execution time of each tool for each program of the benchmark. On average, Wasmati is 9.1x faster than Wassail, demonstrating that our framework outperforms a simpler static analysis tool for WebAssembly.

Note that Wassail is not a vulnerability scanning tool. Wassail was designed only to generate summaries that describe where the information can flow within functions. It offers no functionality that allows us to search for vulnerabilities in \wasm binaries. For this reason, we cannot use it as a baseline for evaluating Wasmati's effectiveness in detecting the vulnerabilities analyzed in Table~\ref{table:query_report}.

\subsection{Query Execution Performance}

Lastly, we measure the query execution time over the SPEC CPU 2017 benchmark to assess its scalability. Table~\ref{tab:queryExecTime-summary} shows the execution times (in seconds) of the WQL queries described in Table~\ref{table:query_report} following the same numbering system. The first six queries are all similar in complexity, which translates into relatively similar execution times for the same binary. 
On average, a binary took less than 77 seconds to run all ten queries, totaling an execution time of around 22 minutes for all binaries. The larger binary, with 3.4M nodes and 44.1M edges, took about 8 minutes to complete all queries.

\begin{table}
\centering
\footnotesize
\tabcolsep=0.1cm
\begin{tabular}{lrrrrrrrrrrrr} 
\hline
\textbf{Query} & \textbf{1} & \textbf{2} & \textbf{3} & \textbf{4} & \textbf{5} & \textbf{6} & \textbf{7} & \textbf{8} & \textbf{9} & \textbf{10} & \textbf{Total} \\
\hline
\textbf{Average} & 4.2 & 10.2 & 174.2 & 1.3 & 1.4 & 1.0 & 0.1 & 5.5 & 1.2 & 2.0 & 201.4 \\
\textbf{Total} & 90.8 & 105.7 & 93.4 & 93.8 & 102.1 & 101.4 & 2.6 & 490.8 & 94.2 & 129.9 & 1304.5 \\
\hline
\end{tabular}
\caption{\small Overview of WQL execution time for SPEC binaries (in seconds). The complete results can be found in the appendix.}
\label{tab:queryExecTime-summary}
\vspace{-0.4cm}
\end{table}

\begin{figure}[t]
    \centering
    \includegraphics[width=0.7\columnwidth]{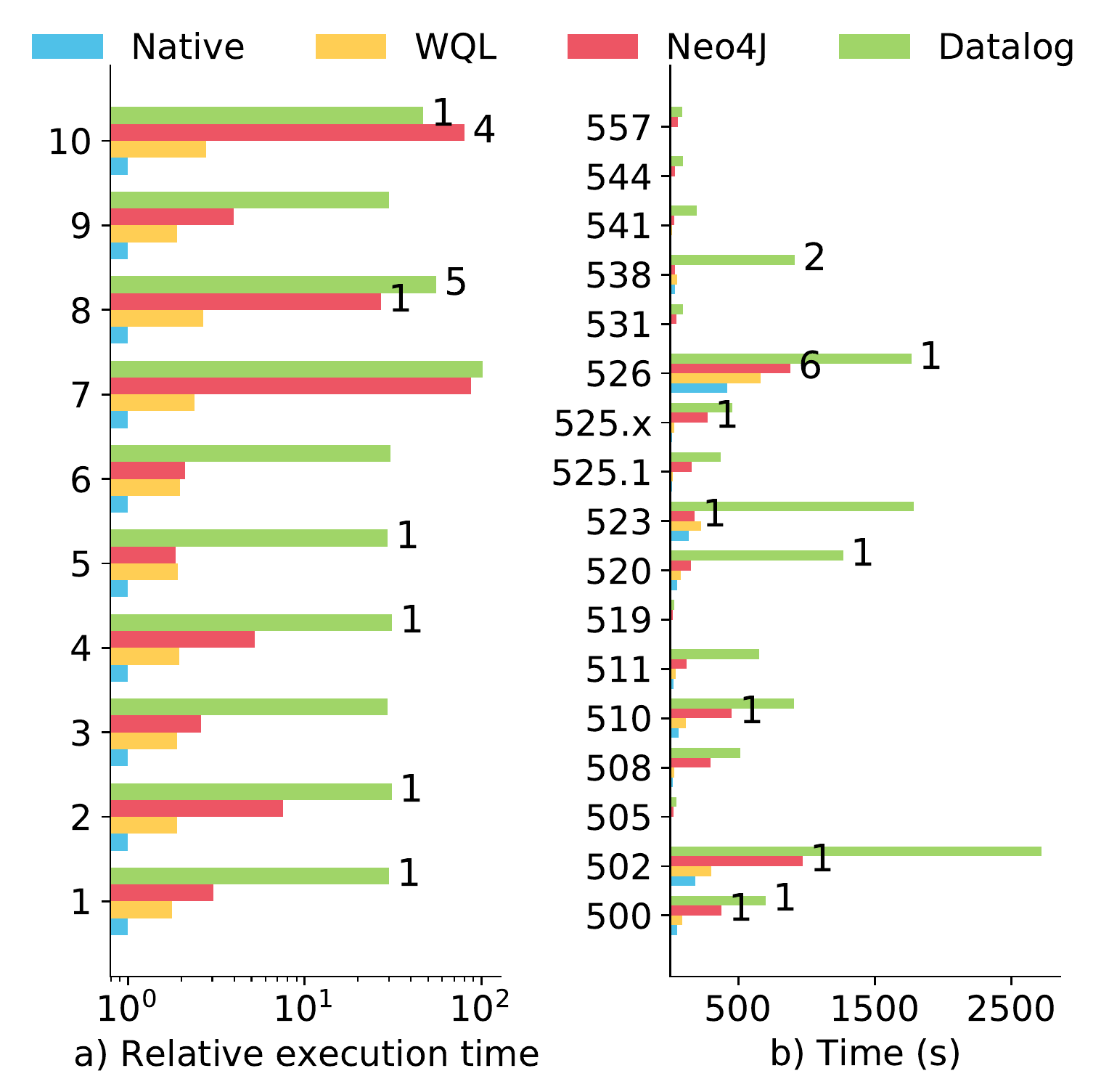}
    \vspace{-0.3cm}
    \caption{\small Execution time comparison across back-ends; a) shows the relative average time for each query over the SPEC dataset. The x-axis is in log scale; native is 1, and all other numbers are relative to native (lower is better); b) shows the total execution times, in seconds, of running all 10 queries in the different SPEC binaries (y-axis). Beside each bar is the number of timeouts that occurred for a query (a), or during the execution of all queries for a binary (b).}
    \label{fig:wasmati_exec}
    \vspace{-0.5cm}
\end{figure}

Figure~\ref{fig:wasmati_exec} compares the execution times between the query back-ends. Neo4j and Datalog were not capable of executing all queries for the SPEC dataset before an established timeout of 10 minutes. In these cases, we capped the query execution time to the 10-minute mark, and proceeded to calculate the described averages.

As it is clear from Figure~\ref{fig:wasmati_exec} a), the native back-end outperforms all other back-ends, but at the cost of a harder query specification process.
Datalog shows consistently high execution times, with the worse result being a 100$\times$ increase in query 7 when compared to the native back-end.
Neo4j shows reasonable performance for simpler queries, such as queries 1 to 6, but quickly explodes in more complex queries, some of which did not finish before a timeout occurred.
Finally, WQL shows significantly better performance than Neo4j and Datalog for most queries. Compared to native execution times, WQL shows, in the worst case, a 3$\times$ overhead, but the queries are much simpler to express and the execution times are still practical.

In Figure~\ref{fig:wasmati_exec} b), six binaries show visible spikes in the execution times. These cases also include timeouts for Datalog or Neo4j. These are also the largest binaries of the SPEC dataset, which means that CPG analysis performance is directly related to the size of the graph, and Neo4j and Datalog are particularly less efficient at analyzing larger CPGs than the native and WQL back-ends.

\section{Limitations}
\label{sec:limitations}

The precision of Wasmati's results is bounded by the queries executed. As such, the ten queries discussed in Section~\ref{sec:queries} may not encode all the patterns that match a specific vulnerability type. For example, a taint vulnerability seizes to exist if proper sanitization is employed, but our taint queries do not check if common sanitization functions are called on the tainted data.

Wasmati can only analyze individual Wasm binaries. 
To uncover actual exploitable vulnerabilities in real applications, the analysis must also consider the application components that interact with the \wasm binary. An interesting direction for future work is to extend Wasmati to model JavaScript calls to exposed functions of a \wasm binary, as well as model accesses to the JavaScript array buffer used as the binary's linear memory.

\section{Related Work}
\label{sec:related}
\mypara{Empirical studies on \wasm:}
Some relevant studies characterize the performance of \wasm engines~\cite{powers2017browsix} as well as the prevalence~\cite{crypto} and security~\cite{lehmann2020everything, hilbig2021empirical} of \wasm code in the wild.
The latter studies on security are of special interest to us, giving a comprehensive account of \wasm security vulnerabilities and how they can be exploited~\cite{lehmann2020everything} and providing a dataset of real-world binaries that we use in our evaluation~\cite{hilbig2021empirical}.

\mypara{Program analyzes for \wasm:}
Since the proposal of the \wasm standard~\cite{WebAssemblyCoreSpecification}, several program analyzes have been designed for tackling the specificities of the language. 
Haas et al.\cite{wasm} proposed a small-step operational semantics for \wasm together with a type system for checking the safety of stack-based operations. Later, Watt~\cite{watt2018mechanising} mechanized both the semantics and the type system introduced in~\cite{wasm}. 
The proof infrastructure of~\cite{watt2018mechanising} was then re-used to formalize and prove the soundness of CT-Wasm~\cite{ct-wasm}, a type-driven extension of WebAssembly for provably secure implementation of cryptographic algorithms. 
More recently, the authors of \cite{program-logic-for-wasm} introduced Wasm Logic, a program logic for modular reasoning about heap-manipulating \wasm programs.
So far, most practical tools for \wasm analysis are based on dynamic analysis. 
Wasabi~\cite{wasabi} is a general framework for instrumenting Wasm binaries and can be used to implement different types of dynamic analyzes.
To the best of our knowledge, there are only three taint analysis tools for \wasm: TaintAssembly~\cite{fu2018taintassembly}, the tool presented in~\cite{szanto2018taint}, and Wassail~\cite{stievenart2020compositional}, with the former two being dynamic and the latter static. Wassail implements a data flow analysis algorithm that has not been tailored for vulnerability detection.
In particular, unlike Wassail, Wasmati specifically tracks constant and function dependencies, which are fundamental to reason about a large number of security vulnerabilities. Importantly, the authors of~\cite{stievenart2020compositional} do not demonstrate how Wassail can be used to enable a vulnerability detection tool. \wasmati is the first CPG-based framework for detecting vulnerabilities in \wasm code.

\mypara{Code Property Graphs:} 
CPGs have been applied to find SQL injection, XSS, and CSRF vulnerabilities in PHP applications \cite{backes2017efficient,pellegrino2017deemon} and, most recently, for detecting CSRF vulnerabilities in client-side JavaScript~\cite{khodayari2021jaw}, all on top of Neo4J's backend. CPGs are at the core of CodeQL~\cite{codeql}, a commercial tool for detecting security vulnerabilities in multiple programming languages, but not \wasm.
Joern~\cite{joern} leverages CPGs to help developers detect vulnerabilities in C/C++ code. Although there is an open-source version of this tool, we did not use it for building Wasmati for three reasons: i) it is a very complex project, ii) Joern's CPGs are generated from an intermediate code representation which hampers their portability and testing on various query back-ends, and iii) developing a custom DSL in C++ helps reduce the overheads of running the queries inside Joern, which runs on top of Scala, a JVM target language.

\section{Conclusion}
\label{sec:conclusion}

In this paper, we presented Wasmati, a static analysis tool for finding vulnerabilities in WebAssembly. It employs optimized techniques for generating CPGs for complex WebAssembly code, as well as four distinct back-ends for query execution. We implemented ten queries for each of the four execution back-ends, capturing different vulnerability types, and extensively tested Wasmati with four heterogeneous datasets.  
Wasmati can scale to large real-world applications and efficiently find vulnerabilities for all tested query types. \textbf{Availability:} Wasmati is publicly available at \url{https://github.com/wasmati/wasmati}.

\section*{Acknowledgements}

This work was supported by Fundação para a Ciência e Tecnologia (FCT) via the SFRH/BD/146698/2019 grant, UIDB/50021/2020 (INESC-ID), project INFOCOS (PTDC/CCI-COM/32378/2017), and by Instituto Superior Técnico, Universidade de Lisboa.

\section*{Biographies}

\noindent \textbf{Tiago Brito} is a Ph.D. student of Information Systems and Computer Engineering at Instituto Superior Técnico (IST) / Universidade de Lisboa and a researcher of the Distributed Systems Group at INESC-ID Lisbon, Portugal. His research focus has been on detecting vulnerabilities in Node.js code using static analysis. He is also a teaching assistant of the digital forensics course at IST.

\noindent \textbf{Pedro Lopes} is a former MSc. student of Information Systems and Computer Engineering at Instituto Superior Técnico / Universidade de Lisboa, which he concluded in 2021, and is now working as a Security Analyst at Siemens, Portugal. His interests consist of vulnerability detection in WebAssembly using static analysis and, most recently, vulnerability research.

\noindent \textbf{Nuno Santos} is an Associate Professor in the Department of Computer Engineering at Instituto Superior Técnico / Universidade de Lisboa and a member of the Distributed Systems Group at INESC-ID Lisbon, Portugal. He received his Ph.D. degree from the Max Planck Institute for Software Systems (MPI-SWS) / Saarland University, Germany, in 2013. His research interests cover multiple topics in systems security, namely trusted computing, web security and vulnerability detection, security policies and GDPR compliance, censorship-resistant and anonymity networks, machine learning applied to traffic analysis, and digital forensics.

\noindent \textbf{José Fragoso Santos} is an Assistant Professor in the Department of Computer Engineering at Instituto Superior Técnico / Universidade de Lisboa, and a research member of the Automated Reasoning and Software Reliability Group at INESC-ID Lisbon. His research interests span the areas of programming languages and software security. He finished his Ph.D. in 2014 at INRIA Sophia Antipolis/University of Nice with a thesis on the topic of information flow control for JavaScript programs. He later joined the Verified Software Group at Imperial College London, where he developed new methods and tools for the verification and testing of Web programs. 


\bibliographystyle{ACM-Reference-Format}
\bibliography{references}


\begin{thebibliography}{43}


\ifx \showCODEN    \undefined \def \showCODEN     #1{\unskip}     \fi
\ifx \showDOI      \undefined \def \showDOI       #1{#1}\fi
\ifx \showISBNx    \undefined \def \showISBNx     #1{\unskip}     \fi
\ifx \showISBNxiii \undefined \def \showISBNxiii  #1{\unskip}     \fi
\ifx \showISSN     \undefined \def \showISSN      #1{\unskip}     \fi
\ifx \showLCCN     \undefined \def \showLCCN      #1{\unskip}     \fi
\ifx \shownote     \undefined \def \shownote      #1{#1}          \fi
\ifx \showarticletitle \undefined \def \showarticletitle #1{#1}   \fi
\ifx \showURL      \undefined \def \showURL       {\relax}        \fi
\providecommand\bibfield[2]{#2}
\providecommand\bibinfo[2]{#2}
\providecommand\natexlab[1]{#1}
\providecommand\showeprint[2][]{arXiv:#2}

\bibitem[\protect\citeauthoryear{??}{cod}{[n.d.]}]%
        {codeql}
 \bibinfo{year}{[n.d.]}\natexlab{}.
\newblock \bibinfo{title}{{CodeQL}}.
\newblock \bibinfo{howpublished}{\url{https://github.com/github/codeql}}.
\newblock
\newblock
\shownote{Accessed: 2021-02-04.}


\bibitem[\protect\citeauthoryear{??}{joe}{[n.d.]}]%
        {joern}
 \bibinfo{year}{[n.d.]}\natexlab{}.
\newblock \bibinfo{title}{{Joern - The Bug Hunter's Workbench}}.
\newblock \bibinfo{howpublished}{\url{https://github.com/joernio/joern}}.
\newblock
\newblock
\shownote{Accessed: 2022-03-16.}


\bibitem[\protect\citeauthoryear{??}{neo}{[n.d.]}]%
        {neo4j}
 \bibinfo{year}{[n.d.]}\natexlab{}.
\newblock \bibinfo{title}{{Neo4j}}.
\newblock \bibinfo{howpublished}{\url{https://neo4j.com/}}.
\newblock
\newblock
\shownote{Accessed: 2021-02-04.}


\bibitem[\protect\citeauthoryear{??}{pol}{[n.d.]}]%
        {polybench}
 \bibinfo{year}{[n.d.]}\natexlab{}.
\newblock \bibinfo{title}{{PolyBenchC: the polyhedral benchmark suite.}}
\newblock
  \bibinfo{howpublished}{\url{http://web.cs.ucla.edu/~pouchet/software/polybench/}}.
\newblock
\newblock
\shownote{Accessed: 2021-02-04.}


\bibitem[\protect\citeauthoryear{??}{shi}{[n.d.]}]%
        {shiftleftSite}
 \bibinfo{year}{[n.d.]}\natexlab{}.
\newblock \bibinfo{title}{{ShiftLeft}}.
\newblock \bibinfo{howpublished}{\url{https://www.shiftleft.io/}}.
\newblock
\newblock
\shownote{Accessed: 2021-02-04.}


\bibitem[\protect\citeauthoryear{??}{plu}{[n.d.]}]%
        {plume}
 \bibinfo{year}{[n.d.]}\natexlab{}.
\newblock \bibinfo{title}{{Soot - A framework for analyzing and transforming
  Java and Android applications}}.
\newblock \bibinfo{howpublished}{\url{https://soot-oss.github.io/soot/}}.
\newblock
\newblock
\shownote{Accessed: 2021-02-04.}


\bibitem[\protect\citeauthoryear{??}{awe}{[n.d.]}]%
        {awesomewasm}
 \bibinfo{year}{[n.d.]}\natexlab{}.
\newblock \bibinfo{title}{{Webassembly Open Source Projects}}.
\newblock
  \bibinfo{howpublished}{\url{https://awesomeopensource.com/projects/webassembly}}.
\newblock
\newblock
\shownote{Accessed: 2021-02-04.}


\bibitem[\protect\citeauthoryear{Backes, Rieck, Skoruppa, Stock, and
  Yamaguchi}{Backes et~al\mbox{.}}{2017}]%
        {backes2017efficient}
\bibfield{author}{\bibinfo{person}{Michael Backes}, \bibinfo{person}{Konrad
  Rieck}, \bibinfo{person}{Malte Skoruppa}, \bibinfo{person}{Ben Stock}, {and}
  \bibinfo{person}{Fabian Yamaguchi}.} \bibinfo{year}{2017}\natexlab{}.
\newblock \showarticletitle{Efficient and flexible discovery of php application
  vulnerabilities}. In \bibinfo{booktitle}{\emph{2017 IEEE european symposium
  on security and privacy (EuroS\&P)}}. IEEE.
\newblock


\bibitem[\protect\citeauthoryear{Clark}{Clark}{2019}]%
        {wasi}
\bibfield{author}{\bibinfo{person}{Lin Clark}.}
  \bibinfo{year}{2019}\natexlab{}.
\newblock \bibinfo{title}{{Standardizing WASI: A system interface to run
  WebAssembly outside the web}}.
\newblock
  \bibinfo{howpublished}{\url{https://hacks.mozilla.org/2019/03/standardizing-wasi-a-webassembly-system-interface/}}.
\newblock
\newblock
\shownote{Accessed: 2021-02-04.}


\bibitem[\protect\citeauthoryear{CVE}{CVE}{[n.d.]}]%
        {CVE-2018-14550}
\bibfield{author}{\bibinfo{person}{CVE}.} \bibinfo{year}{[n.d.]}\natexlab{}.
\newblock \bibinfo{title}{{CVE-2018-14550}}.
\newblock
  \bibinfo{howpublished}{\url{https://www.cvedetails.com/cve/CVE-2018-14550/}}.
\newblock
\newblock
\shownote{Accessed: 2021-02-04.}


\bibitem[\protect\citeauthoryear{CWE}{CWE}{[n.d.]}]%
        {cwe_c}
\bibfield{author}{\bibinfo{person}{CWE}.} \bibinfo{year}{[n.d.]}\natexlab{}.
\newblock \bibinfo{title}{{CWE VIEW: Weaknesses in Software Written in C}}.
\newblock
  \bibinfo{howpublished}{\url{https://cwe.mitre.org/data/definitions/658.html}}.
\newblock
\newblock
\shownote{Accessed: 2021-02-04.}


\bibitem[\protect\citeauthoryear{{Emscripten}}{{Emscripten}}{[n.d.]}]%
        {emscripten}
\bibfield{author}{\bibinfo{person}{{Emscripten}}.}
  \bibinfo{year}{[n.d.]}\natexlab{}.
\newblock \bibinfo{title}{{An Open Source LLVM to JavaScript and WebAssembly
  compiler}}.
\newblock \bibinfo{howpublished}{\url{https://emscripten.org}}.
\newblock
\newblock
\shownote{Accessed: 2021-02-04.}


\bibitem[\protect\citeauthoryear{Foundation}{Foundation}{[n.d.]}]%
        {nodejs}
\bibfield{author}{\bibinfo{person}{OpenJS Foundation}.}
  \bibinfo{year}{[n.d.]}\natexlab{}.
\newblock \bibinfo{title}{{Node.js}}.
\newblock \bibinfo{howpublished}{\url{https://nodejs.org/en/}}.
\newblock
\newblock
\shownote{Accessed: 2021-02-04.}


\bibitem[\protect\citeauthoryear{Fu, Lin, and Inge}{Fu et~al\mbox{.}}{2018}]%
        {fu2018taintassembly}
\bibfield{author}{\bibinfo{person}{William Fu}, \bibinfo{person}{Raymond Lin},
  {and} \bibinfo{person}{Daniel Inge}.} \bibinfo{year}{2018}\natexlab{}.
\newblock \showarticletitle{Taintassembly: Taint-based information flow control
  tracking for webassembly}.
\newblock \bibinfo{journal}{\emph{arXiv preprint}} (\bibinfo{year}{2018}).
\newblock


\bibitem[\protect\citeauthoryear{Group}{Group}{[n.d.]}]%
        {wabt}
\bibfield{author}{\bibinfo{person}{WebAssembly~Community Group}.}
  \bibinfo{year}{[n.d.]}\natexlab{}.
\newblock \bibinfo{title}{{The WebAssembly Binary Toolkit}}.
\newblock \bibinfo{howpublished}{\url{https://github.com/WebAssembly/wabt}}.
\newblock
\newblock
\shownote{Accessed: 2021-02-04.}


\bibitem[\protect\citeauthoryear{Haas, Rossberg, Schuff, Titzer, Holman,
  Gohman, Wagner, Zakai, and Bastien}{Haas et~al\mbox{.}}{2017}]%
        {wasm}
\bibfield{author}{\bibinfo{person}{Andreas Haas}, \bibinfo{person}{Andreas
  Rossberg}, \bibinfo{person}{Derek~L. Schuff}, \bibinfo{person}{Ben~L.
  Titzer}, \bibinfo{person}{Michael Holman}, \bibinfo{person}{Dan Gohman},
  \bibinfo{person}{Luke Wagner}, \bibinfo{person}{Alon Zakai}, {and}
  \bibinfo{person}{JF Bastien}.} \bibinfo{year}{2017}\natexlab{}.
\newblock \showarticletitle{Bringing the Web Up to Speed with WebAssembly}.
\newblock  (\bibinfo{year}{2017}).
\newblock


\bibitem[\protect\citeauthoryear{Hall and Ramachandran}{Hall and
  Ramachandran}{2019}]%
        {wasmiot}
\bibfield{author}{\bibinfo{person}{Adam Hall} {and} \bibinfo{person}{Umakishore
  Ramachandran}.} \bibinfo{year}{2019}\natexlab{}.
\newblock \showarticletitle{An Execution Model for Serverless Functions at the
  Edge}. In \bibinfo{booktitle}{\emph{Proceedings of the International
  Conference on Internet of Things Design and Implementation}}
  \emph{(\bibinfo{series}{IoTDI '19})}. \bibinfo{publisher}{ACM}.
\newblock


\bibitem[\protect\citeauthoryear{Hilbig, Lehmann, and Pradel}{Hilbig
  et~al\mbox{.}}{2021}]%
        {hilbig2021empirical}
\bibfield{author}{\bibinfo{person}{Aaron Hilbig}, \bibinfo{person}{Daniel
  Lehmann}, {and} \bibinfo{person}{Michael Pradel}.}
  \bibinfo{year}{2021}\natexlab{}.
\newblock \showarticletitle{An Empirical Study of Real-World WebAssembly
  Binaries}.
\newblock  (\bibinfo{year}{2021}).
\newblock


\bibitem[\protect\citeauthoryear{{ISO Central Secretary}}{{ISO Central
  Secretary}}{2011}]%
        {iso_iec_9899_2011}
\bibfield{author}{\bibinfo{person}{{ISO Central Secretary}}.}
  \bibinfo{year}{2011}\natexlab{}.
\newblock \bibinfo{booktitle}{}.
\newblock \bibinfo{type}{Standard} ISO/IEC 9899:2011.
  \bibinfo{institution}{International Organization for Standardization}.
\newblock
\urldef\tempurl%
\url{https://www.iso.org/standard/57853.html}
\showURL{%
\tempurl}


\bibitem[\protect\citeauthoryear{Jordan, Scholz, and Suboti{\'c}}{Jordan
  et~al\mbox{.}}{2016}]%
        {jordan2016souffle}
\bibfield{author}{\bibinfo{person}{Herbert Jordan}, \bibinfo{person}{Bernhard
  Scholz}, {and} \bibinfo{person}{Pavle Suboti{\'c}}.}
  \bibinfo{year}{2016}\natexlab{}.
\newblock \showarticletitle{Souffle: On synthesis of program analyzers}. In
  \bibinfo{booktitle}{\emph{International Conference on Computer Aided
  Verification}}.
\newblock


\bibitem[\protect\citeauthoryear{Khodayari and Pellegrino}{Khodayari and
  Pellegrino}{2021}]%
        {khodayari2021jaw}
\bibfield{author}{\bibinfo{person}{Soheil Khodayari} {and}
  \bibinfo{person}{Giancarlo Pellegrino}.} \bibinfo{year}{2021}\natexlab{}.
\newblock \showarticletitle{JAW: Studying Client-side CSRF with Hybrid Property
  Graphs and Declarative Traversals}. In \bibinfo{booktitle}{\emph{USENIX
  Security Symposium}}.
\newblock


\bibitem[\protect\citeauthoryear{Lehmann, Kinder, and Pradel}{Lehmann
  et~al\mbox{.}}{2020}]%
        {lehmann2020everything}
\bibfield{author}{\bibinfo{person}{Daniel Lehmann}, \bibinfo{person}{Johannes
  Kinder}, {and} \bibinfo{person}{Michael Pradel}.}
  \bibinfo{year}{2020}\natexlab{}.
\newblock \showarticletitle{Everything old is new again: Binary security of
  webassembly}. In \bibinfo{booktitle}{\emph{29th USENIX Security Symposium}}.
\newblock


\bibitem[\protect\citeauthoryear{Lehmann and Pradel}{Lehmann and
  Pradel}{2019}]%
        {wasabi}
\bibfield{author}{\bibinfo{person}{Daniel Lehmann} {and}
  \bibinfo{person}{Michael Pradel}.} \bibinfo{year}{2019}\natexlab{}.
\newblock \showarticletitle{Wasabi}.
\newblock \bibinfo{journal}{\emph{Proceedings of the Twenty-Fourth
  International Conference on Architectural Support for Programming Languages
  and Operating Systems - ASPLOS ’19}} (\bibinfo{year}{2019}).
\newblock


\bibitem[\protect\citeauthoryear{McFadden, Lukasiewicz, Dileo, and
  Engler}{McFadden et~al\mbox{.}}{2018}]%
        {blackhat}
\bibfield{author}{\bibinfo{person}{Brian McFadden}, \bibinfo{person}{Tyler
  Lukasiewicz}, \bibinfo{person}{Jeff Dileo}, {and} \bibinfo{person}{Justin
  Engler}.} \bibinfo{year}{2018}\natexlab{}.
\newblock \showarticletitle{Security Chasms of WASM}.
\newblock \bibinfo{journal}{\emph{BlackHat US-18}} (\bibinfo{year}{2018}).
\newblock
\urldef\tempurl%
\url{https://i.blackhat.com/us-18/Thu-August-9/us-18-Lukasiewicz-WebAssembly-A-New-World-of-Native_Exploits-On-The-Web-wp.pdf}
\showURL{%
\tempurl}


\bibitem[\protect\citeauthoryear{{Mitre}}{{Mitre}}{[n.d.]}]%
        {CWE-242}
\bibfield{author}{\bibinfo{person}{{Mitre}}.}
  \bibinfo{year}{[n.d.]}\natexlab{}.
\newblock \bibinfo{title}{{CWE-242: Use of Inherently Dangerous Function}}.
\newblock
  \bibinfo{howpublished}{\url{https://cwe.mitre.org/data/definitions/242}}.
\newblock
\newblock
\shownote{Accessed: 2021-02-04.}


\bibitem[\protect\citeauthoryear{Musch, Wressnegger, Johns, and Rieck}{Musch
  et~al\mbox{.}}{2019}]%
        {crypto}
\bibfield{author}{\bibinfo{person}{Marius Musch}, \bibinfo{person}{Christian
  Wressnegger}, \bibinfo{person}{Martin Johns}, {and} \bibinfo{person}{Konrad
  Rieck}.} \bibinfo{year}{2019}\natexlab{}.
\newblock \showarticletitle{New Kid on the Web: A Study on the Prevalence of
  WebAssembly in the Wild}. In \bibinfo{booktitle}{\emph{International
  Conference on Detection of Intrusions and Malware, and Vulnerability
  Assessment}}. Springer.
\newblock


\bibitem[\protect\citeauthoryear{Møller and Schwartzbach}{Møller and
  Schwartzbach}{2020}]%
        {moller:book:2020}
\bibfield{author}{\bibinfo{person}{Anders Møller} {and}
  \bibinfo{person}{Michael~I. Schwartzbach}.} \bibinfo{year}{2020}\natexlab{}.
\newblock \bibinfo{booktitle}{\emph{Static Program Analysis}}.
\newblock \bibinfo{publisher}{Aarhus University}.
\newblock


\bibitem[\protect\citeauthoryear{{Parity}}{{Parity}}{[n.d.]}]%
        {smartcontracts}
\bibfield{author}{\bibinfo{person}{{Parity}}.}
  \bibinfo{year}{[n.d.]}\natexlab{}.
\newblock \bibinfo{title}{{Blockchain Infrastructure for the Decentralised
  Web}}.
\newblock \bibinfo{howpublished}{\url{https://www.parity.io}}.
\newblock
\newblock
\shownote{Accessed: 2021-02-04.}


\bibitem[\protect\citeauthoryear{{Pat Hickey}}{{Pat Hickey}}{[n.d.]}]%
        {lucet}
\bibfield{author}{\bibinfo{person}{{Pat Hickey}}.}
  \bibinfo{year}{[n.d.]}\natexlab{}.
\newblock \bibinfo{title}{{Announcing Lucet: Fastly’s native WebAssembly
  compiler and runtime}}.
\newblock
  \bibinfo{howpublished}{\url{https://www.fastly.com/blog/announcing-lucet-fastly-native-webassembly-compiler-runtime}}.
\newblock
\newblock
\shownote{Accessed: 2021-02-04.}


\bibitem[\protect\citeauthoryear{Pellegrino, Johns, Koch, Backes, and
  Rossow}{Pellegrino et~al\mbox{.}}{2017}]%
        {pellegrino2017deemon}
\bibfield{author}{\bibinfo{person}{Giancarlo Pellegrino},
  \bibinfo{person}{Martin Johns}, \bibinfo{person}{Simon Koch},
  \bibinfo{person}{Michael Backes}, {and} \bibinfo{person}{Christian Rossow}.}
  \bibinfo{year}{2017}\natexlab{}.
\newblock \showarticletitle{Deemon: Detecting CSRF with dynamic analysis and
  property graphs}. In \bibinfo{booktitle}{\emph{Proceedings of the 2017 ACM
  SIGSAC Conference on Computer and Communications Security}}.
\newblock


\bibitem[\protect\citeauthoryear{Powers, Vilk, and Berger}{Powers
  et~al\mbox{.}}{2017}]%
        {powers2017browsix}
\bibfield{author}{\bibinfo{person}{Bobby Powers}, \bibinfo{person}{John Vilk},
  {and} \bibinfo{person}{Emery~D Berger}.} \bibinfo{year}{2017}\natexlab{}.
\newblock \showarticletitle{Browsix: Bridging the gap between unix and the
  browser}.
\newblock \bibinfo{journal}{\emph{ACM SIGPLAN Notices}} (\bibinfo{year}{2017}).
\newblock


\bibitem[\protect\citeauthoryear{Rossberg}{Rossberg}{[n.d.]}]%
        {WebAssemblyCoreSpecification}
\bibfield{author}{\bibinfo{person}{Andreas Rossberg}.}
  \bibinfo{year}{[n.d.]}\natexlab{}.
\newblock \bibinfo{title}{{WebAssembly Core Specification}}.
\newblock
\newblock
\urldef\tempurl%
\url{https://www.w3.org/TR/wasm-core-1/}
\showURL{%
\tempurl}
\newblock
\shownote{\url{https://webassembly.github.io/spec/core/_download/WebAssembly.pdf}.}


\bibitem[\protect\citeauthoryear{ShiftLeft}{ShiftLeft}{[n.d.]}]%
        {shiftleft}
\bibfield{author}{\bibinfo{person}{ShiftLeft}.}
  \bibinfo{year}{[n.d.]}\natexlab{}.
\newblock \bibinfo{title}{{Analyze Your Code in Less than 10 Minutes}}.
\newblock
  \bibinfo{howpublished}{\url{https://www.youtube.com/watch?v=VaxSi1yC9mo}}.
\newblock
\newblock
\shownote{Accessed: 2021-02-04.}


\bibitem[\protect\citeauthoryear{{SPEC}}{{SPEC}}{[n.d.]}]%
        {spec2017}
\bibfield{author}{\bibinfo{person}{{SPEC}}.} \bibinfo{year}{[n.d.]}\natexlab{}.
\newblock \bibinfo{title}{{SPEC CPU® 2017}}.
\newblock \bibinfo{howpublished}{\url{https://www.spec.org/cpu2017/}}.
\newblock
\newblock
\shownote{Accessed: 2021-02-04.}


\bibitem[\protect\citeauthoryear{Sti{\'e}venart and De~Roover}{Sti{\'e}venart
  and De~Roover}{2020}]%
        {stievenart2020compositional}
\bibfield{author}{\bibinfo{person}{Quentin Sti{\'e}venart} {and}
  \bibinfo{person}{Coen De~Roover}.} \bibinfo{year}{2020}\natexlab{}.
\newblock \showarticletitle{Compositional Information Flow Analysis for
  WebAssembly Programs}. In \bibinfo{booktitle}{\emph{2020 IEEE 20th
  International Working Conference on Source Code Analysis and Manipulation
  (SCAM)}}. IEEE.
\newblock


\bibitem[\protect\citeauthoryear{{STT}}{{STT}}{[n.d.]}]%
        {stt}
\bibfield{author}{\bibinfo{person}{{STT}}.} \bibinfo{year}{[n.d.]}\natexlab{}.
\newblock \bibinfo{title}{{We are STT}}.
\newblock \bibinfo{howpublished}{\url{https://sectt.github.io/}}.
\newblock
\newblock
\shownote{Accessed: 2021-02-04.}


\bibitem[\protect\citeauthoryear{Szanto, Tamm, and Pagnoni}{Szanto
  et~al\mbox{.}}{2018}]%
        {szanto2018taint}
\bibfield{author}{\bibinfo{person}{Aron Szanto}, \bibinfo{person}{Timothy
  Tamm}, {and} \bibinfo{person}{Artidoro Pagnoni}.}
  \bibinfo{year}{2018}\natexlab{}.
\newblock \showarticletitle{Taint tracking for webassembly}.
\newblock \bibinfo{journal}{\emph{arXiv preprint}} (\bibinfo{year}{2018}).
\newblock


\bibitem[\protect\citeauthoryear{Wasmer}{Wasmer}{[n.d.]}]%
        {wasmer}
\bibfield{author}{\bibinfo{person}{Wasmer}.} \bibinfo{year}{[n.d.]}\natexlab{}.
\newblock \bibinfo{title}{{An open-source runtime for executing WebAssembly on
  the Server}}.
\newblock \bibinfo{howpublished}{\url{https://wasmer.io/}}.
\newblock
\newblock
\shownote{Accessed: 2021-02-04.}


\bibitem[\protect\citeauthoryear{Watt}{Watt}{2018}]%
        {watt2018mechanising}
\bibfield{author}{\bibinfo{person}{Conrad Watt}.}
  \bibinfo{year}{2018}\natexlab{}.
\newblock \showarticletitle{Mechanising and verifying the WebAssembly
  specification}. In \bibinfo{booktitle}{\emph{Proceedings of the 7th ACM
  SIGPLAN International Conference on Certified Programs and Proofs}}.
\newblock


\bibitem[\protect\citeauthoryear{Watt, Maksimovic, Krishnaswami, and
  Gardner}{Watt et~al\mbox{.}}{2019a}]%
        {program-logic-for-wasm}
\bibfield{author}{\bibinfo{person}{Conrad Watt}, \bibinfo{person}{Petar
  Maksimovic}, \bibinfo{person}{Neelakantan~R. Krishnaswami}, {and}
  \bibinfo{person}{Philippa Gardner}.} \bibinfo{year}{2019}\natexlab{a}.
\newblock \showarticletitle{A Program Logic for First-Order Encapsulated
  {WebAssembly}}. In \bibinfo{booktitle}{\emph{33rd European Conference on
  Object-Oriented Programming (ECOOP 2019)}},
  \bibfield{editor}{\bibinfo{person}{Alastair~F. Donaldson}} (Ed.).
\newblock


\bibitem[\protect\citeauthoryear{Watt, Renner, Popescu, Cauligi, and
  Stefan}{Watt et~al\mbox{.}}{2019b}]%
        {ct-wasm}
\bibfield{author}{\bibinfo{person}{Conrad Watt}, \bibinfo{person}{John Renner},
  \bibinfo{person}{Natalie Popescu}, \bibinfo{person}{Sunjay Cauligi}, {and}
  \bibinfo{person}{Deian Stefan}.} \bibinfo{year}{2019}\natexlab{b}.
\newblock \showarticletitle{CT-wasm: Type-driven Secure Cryptography for the
  Web Ecosystem}.
\newblock \bibinfo{journal}{\emph{Proc. ACM Program. Lang.}}
  (\bibinfo{year}{2019}).
\newblock


\bibitem[\protect\citeauthoryear{Yamaguchi, Golde, Arp, and Rieck}{Yamaguchi
  et~al\mbox{.}}{2014}]%
        {cpg}
\bibfield{author}{\bibinfo{person}{Fabian Yamaguchi}, \bibinfo{person}{Nico
  Golde}, \bibinfo{person}{Daniel Arp}, {and} \bibinfo{person}{Konrad Rieck}.}
  \bibinfo{year}{2014}\natexlab{}.
\newblock \showarticletitle{Modeling and Discovering Vulnerabilities with Code
  Property Graphs}.
\newblock \bibinfo{journal}{\emph{Proceedings - IEEE Symposium on Security and
  Privacy}} (\bibinfo{year}{2014}).
\newblock


\bibitem[\protect\citeauthoryear{Yamaguchi, Maier, Gascon, and Rieck}{Yamaguchi
  et~al\mbox{.}}{2015}]%
        {yamaguchi2015automatic}
\bibfield{author}{\bibinfo{person}{Fabian Yamaguchi}, \bibinfo{person}{Alwin
  Maier}, \bibinfo{person}{Hugo Gascon}, {and} \bibinfo{person}{Konrad Rieck}.}
  \bibinfo{year}{2015}\natexlab{}.
\newblock \showarticletitle{Automatic inference of search patterns for
  taint-style vulnerabilities}. In \bibinfo{booktitle}{\emph{2015 IEEE
  Symposium on Security and Privacy}}. IEEE.
\newblock


\end{thebibliography}

\appendix

\clearpage
\onecolumn
\section{Appendix}

\subsection{Code Property Graph}
\subsubsection{Node and Edge definitions}
\renewcommand{\arraystretch}{1.2}
\begin{table}[h]
\begin{subtable}[b]{.22\linewidth}
\centering
\begin{tabular}{|c|c|}
\hline
\textbf{Property}& \textbf{Value}\\\hline\hline
$type$ & AST \\\hline
\end{tabular}
\caption{AST edge.}
\end{subtable}
\hfill
\begin{subtable}[b]{.22\linewidth}
\centering
\begin{tabular}{|c|c|}
\hline
\textbf{Property}& \textbf{Value}\\\hline\hline
    $type$ & CG \\\hline
\end{tabular}
\caption{CG edge.}
\end{subtable}
\hfill
\begin{subtable}[b]{.22\linewidth}
\centering
\begin{tabular}{|c|c|}
\hline
\textbf{Property}& \textbf{Value}\\\hline\hline
    $type$ & CFG \\\hline
    $label$ & $\Sigma$ \\\hline
\end{tabular}
\caption{CFG edge.}
\end{subtable}
\hfill
\begin{subtable}[b]{.22\linewidth}
\centering
\begin{tabular}{|c|c|}
\hline
\textbf{Property}& \textbf{Value}\\\hline\hline
     $type$ & DDG \\ \hline
     $label$ & $\Sigma_P$ \\ \hline
     $ddgType$ & $\Sigma_T$ \\ \hdashline \hdashline
     $valueType$ &  $\Sigma_V$\\ \hline
     $value$ &  $\mathbb{R}$\\ \hline
\end{tabular}
\caption{DDG edge.}
\end{subtable}
\caption{Edges.}
\end{table}

\begin{minipage}{0.48\textwidth}
$\Sigma =$ \{false, true, default\} $\cup$ $\mathbb{N}_0$~\\
$\Sigma_P =$ $\Sigma \cup FunctionName$
\end{minipage}
\begin{minipage}{0.48\textwidth}
$\Sigma_T =$ \{Global, Local, Const, Control, Function\}\\
$\Sigma_V =$ \{i32, i64, f32, f64\} 
\end{minipage}

\dotfill

\begin{table}[h]
\parbox[b]{.27\linewidth}{
\centering
\centering
\begin{tabular}{|c|c|}
\hline
\textbf{Property}& \textbf{Value}\\\hline\hline
$id$ & $\mathbb{N}_0$ \\\hline
$type$ & Module \\\hline
$name$ & $\{string\}$ \\\hline
$index$ & $\mathbb{N}_0$ \\\hline
$nargs$ & $\mathbb{N}_0$ \\\hline
$nlocals$ & $\mathbb{N}_0$ \\\hline
$nresults$ & $\mathbb{N}_0$ \\\hline
$isImport$ & \{false,true\} \\\hline
$isExport$ & \{false,true\} \\\hline
\end{tabular}
\caption{Function node.}
}
\hfill
\parbox[b]{.35\linewidth}{
\centering
\begin{tabular}[t]{|c|c|}
\hline
\textbf{Property}& \textbf{Value}\\\hline\hline
$id$ & $\mathbb{N}_0$ \\\hline
$type$ & Module \\\hline
$name$ & $\{string\}$ \\\hline
\end{tabular}
\caption{Module node.}
~\\
\begin{tabular}[b]{|c|c|}
\hline
\textbf{Property}& \textbf{Value}\\\hline\hline
$id$ & $\mathbb{N}_0$ \\\hline
$type$ & Instruction \\ \hline
$instType$ & $\Sigma_{S}$ \\\hline
\end{tabular}
\caption{Simple instruction node.}
\label{table:simpleInst}
}
\hfill
\parbox[b]{.35\linewidth}{
\centering
\begin{tabular}{|c|c|}
\hline
\textbf{Property}& \textbf{Value}\\\hline\hline
$id$ & $\mathbb{N}_0$ \\\hline
$type$ & $\Sigma_T$ \\\hline
\end{tabular}
\caption{Simple node.}
~\\
\begin{tabular}{|c|c|}
\hline
\textbf{Property}& \textbf{Value}\\\hline\hline
$id$ & $\mathbb{N}_0$ \\\hline
$type$ & Instruction \\ \hline
$instType$ & Const \\\hline
$valueType$ & \{i32, i64, f32, f64\} \\\hline
$value$ & $\mathbb{R}$ \\\hline
\end{tabular}
\caption{Constant node.}
}
\end{table}
\noindent
$\Sigma_{T} =$ \{FunctionSignature, Parameters, Locals, Results, Else, Trap, Start\}.
~\\
$\Sigma_{S} =$ \{Nop, Unreachable, Return, BrTable, Drop, Select, MemorySize, MemoryGrow, CallIndirect\}.

\dotfill

\begin{table}[h]
\parbox[b]{.3\linewidth}{
\centering
\begin{tabular}{|c|c|}
\hline
\textbf{Property}& \textbf{Value}\\\hline\hline
$id$ & $\mathbb{N}_0$ \\\hline
$type$ & Instruction \\ \hline
$instType$ & $\Sigma_T$ \\\hline
$label$ & $\{string\}$ \\\hline
\end{tabular}
\caption{Labelled node.}
}
\hfill
\parbox[b]{.3\linewidth}{
\centering
\begin{tabular}{|c|c|}
\hline
\textbf{Property}& \textbf{Value}\\\hline\hline
$id$ & $\mathbb{N}_0$ \\\hline
$type$ & Instruction \\ \hline
$instType$ & Block \\\hline
$label$ & $\{string\}$ \\\hline
$nresults$ & $\mathbb{N}_0$ \\\hline
\end{tabular}
\caption{Block node.}
}
\hfill
\parbox[b]{.3\linewidth}{
\centering
\begin{tabular}{|c|c|}
\hline
\textbf{Property}& \textbf{Value}\\\hline\hline
$id$ & $\mathbb{N}_0$ \\\hline
$type$ & Instruction \\ \hline
$instType$ & If \\\hline
$label$ & $\{string\}$ \\\hline
$hasElse$ & \{false,true\} \\\hline
\end{tabular}
\caption{If node.}
}
\end{table}

\noindent
$\Sigma_T =$ \{Br, BrIf, GlobalGet, GlobalSet, LocalGet, LocalSet, LocalTee, Call, BeginBlock\}.

\begin{table}[h!]
\begin{subtable}[b]{.25\linewidth}
\centering
\begin{tabular}{|c|c|}
\hline
\textbf{Property}& \textbf{Value}\\\hline\hline
$id$ & $\mathbb{N}_0$ \\\hline
$type$ & Instruction \\\hline
$instType$ & Binary \\\hline
$opcode$ & $\{binop\}$ \\\hline
\end{tabular}
\caption{Binary node.}
\end{subtable}
\hfill
\begin{subtable}[b]{.25\linewidth}
\centering
\begin{tabular}{|c|c|}
\hline
\textbf{Property}& \textbf{Value}\\\hline\hline
$id$ & $\mathbb{N}_0$ \\\hline
$type$ & Instruction \\\hline
$instType$ & Compare \\\hline
$opcode$ & $\{relop\}$ \\\hline
\end{tabular}
\caption{Compare node.}
\end{subtable}
\hfill
\parbox[b]{.25\linewidth}{
\centering
\begin{tabular}{|c|c|}
\hline
\textbf{Property}& \textbf{Value}\\\hline\hline
$id$ & $\mathbb{N}_0$ \\\hline
$type$ & Instruction \\\hline
$instType$ & Load \\\hline
$offset$ & $\mathbb{N}_0$  \\\hline
\end{tabular}
\caption{Load node.}
}
\hfill
\begin{subtable}[b]{.25\linewidth}
\centering
\begin{tabular}{|c|c|}
\hline
\textbf{Property}& \textbf{Value}\\\hline\hline
$id$ & $\mathbb{N}_0$ \\\hline
$type$ & Instruction \\\hline
$instType$ & Convert \\\hline
$opcode$ & $\{cutop\}$ \\\hline
\end{tabular}
\caption{Convert node.}
\end{subtable}
\hfill
\begin{subtable}[b]{.25\linewidth}
\centering
\begin{tabular}{|c|c|}
\hline
\textbf{Property}& \textbf{Value}\\\hline\hline
$id$ & $\mathbb{N}_0$ \\\hline
$type$ & Instruction \\\hline
$instType$ & Unary \\\hline
$opcode$ & $\{unop\}$ \\\hline
\end{tabular}
\caption{Unary node.}
\end{subtable}
\hfill
\parbox[b]{.25\linewidth}{
\centering
\begin{tabular}{|c|c|}
\hline
\textbf{Property}& \textbf{Value}\\\hline\hline
$id$ & $\mathbb{N}_0$ \\\hline
$type$ & Instruction \\\hline
$instType$ & Store \\\hline
$offset$ & $\mathbb{N}_0$  \\\hline
\end{tabular}
\caption{Store node.}
}
\captionsetup{singlelinecheck=false,justification=raggedright, margin=0.15\linewidth}
\caption{Opcode instructions.}
\end{table}

\begin{algorithm}[t]
\small
\SetAlgoLined
\SetKwComment{Comment}{}{}
\KwIn{Sequencial Instructions}
\KwResult{Folded Instructions}
 $ST := \emptyset$  \Comment*[l]{\quad$\triangleright$ Stack for instructions (LIFO)}
 $RS := \emptyset$  \Comment*[l]{\quad$\triangleright$ List for instructions (FIFO)}
\ForEach(){inst  $\in$ Instructions}{
    ($nargs,nresults$) $:=$ GetArity($inst$)\;
    \ForEach(){$i \in \{1...nargs\}$} {
        $child$ $:=$ Pop($ST$) \;
        AddEdge($inst$.id, $child$.id, ``AST'') \;
    }
    \uIf{$nresults = 0$}{ Append($inst$, $RS$) }
    \Else{ Push($inst$, $ST$)}
 }
 \Return RS \;
 \caption{\small AST Folding.}
 \label{algorithm:ast_folding}
\end{algorithm}

\FloatBarrier

\subsubsection{Transfer Functions}
\begin{table*}[h!]
\centering
\small
\begin{tabular}{rcll}
\hline
$\mathcal{T}($\verb|t.const c|, $(\gsto,\lsto,\stack))$ & $\hookrightarrow_{\id}$ & $(\gsto,\lsto,\stack::\constdep{\id}{c, t})$\\
$\mathcal{T}($\verb|t.unop|, $(\gsto,\lsto,\stack::v_0))$ & $\hookrightarrow_{\id}$ & $(\gsto,\lsto,\stack::v_0)$ \\
$\mathcal{T}($\verb|t.binop|, $(\gsto,\lsto,\stack::v_0::v_1))$ & $\hookrightarrow_{\id}$ & $(\gsto,\lsto,\stack::v_0\cup v_1))$&\\
$\mathcal{T}($\verb|t.relop|, $(\gsto,\lsto,\stack::v_0::v_1))$ & $\hookrightarrow_{\id}$  & $(\gsto,\lsto,\stack::v_0\cup v_1)$&\\
$\mathcal{T}($\verb|t|$_2$\verb|.cvtop_t|$_1$\verb|_sx|, $(\gsto,\lsto,\stack::v_0))$ & $\hookrightarrow_{\id}$  & $(\gsto,\lsto,\stack::v_0)$&\\
$\mathcal{T}($\verb|drop|, $(\gsto,\lsto,\stack::v_0))$ & $\hookrightarrow_{\id}$  & $(\gsto,\lsto,\stack)$ &\\
$\mathcal{T}($\verb|select|, $(\gsto,\lsto,\stack::v_0::v_1::v_2))$ & $\hookrightarrow_{\id}$  & $(\gsto,\lsto,\stack::v_0\cup v_1)$ &\\
$\mathcal{T}($\verb|local.get x|, $(\gsto,\lsto,\stack))$ & $\hookrightarrow_{\id}$  & $(\gsto,\lsto,\stack::\lsto(f,\tvar{x}))$ &\\
$\mathcal{T}($\verb|local.set x|, $(\gsto,\lsto,\stack::v_0))$ & $\hookrightarrow_{\id}$ & $(\gsto,\lsto[\tvar{x} \mapsto v_0],\stack)$ &\\
$\mathcal{T}($\verb|local.tee x|, $(\gsto,\lsto,\stack::v_0))$ & $\hookrightarrow_{\id}$ & $(\gsto,\lsto[\tvar{x} \mapsto v_0],\stack::v_0)$ &\\
$\mathcal{T}($\verb|global.get x|, $(\gsto,\lsto,\stack))$ & $\hookrightarrow_{\id}$ & $(\gsto,\lsto,\stack::\gsto(\text{x}))$ &\\
$\mathcal{T}($\verb|global.set x|, $(\gsto,\lsto,\stack::v_0))$ & $\hookrightarrow_{\id}$ & $(\gsto[\tvar{x} \mapsto v_0],\lsto,\stack)$ &\\
$\mathcal{T}($\verb|t.load|, $(\gsto,\lsto,\stack::v_0))$ & $\hookrightarrow_{\id}$ & $(\gsto,\lsto,\stack)$ &\\
$\mathcal{T}($\verb|t.store|, $(\gsto,\lsto,\stack::v_0::v_1))$ & $\hookrightarrow_{\id}$ & $(\gsto,\lsto,\stack)$ &\\
$\mathcal{T}($\verb|memory.size|, $(\gsto,\lsto,\stack))$ & $\hookrightarrow_{\id}$ & $(\gsto,\lsto,\stack::\{\})$ &\\
$\mathcal{T}($\verb|memory.grow|, $(\gsto,\lsto,\stack::v_0))$ & $\hookrightarrow_{\id}$ & $(\gsto,\lsto,\stack::\{\})$ &\\
$\mathcal{T}($\{\verb|nop| $|$ \verb|br b| $|$ \verb|return|\}, $(\gsto,\lsto,\stack))$ & $\hookrightarrow_{\id}$ & $(\gsto,\lsto,\stack)$ &\\
$\mathcal{T}($\verb|unreachable|, $(\gsto,\lsto,\stack))$ & $\hookrightarrow_{\id}$ & $(\gsto,\lsto,\stack)$ &\\
$\mathcal{T}($\verb|beginBlock b|, $(\gsto,\lsto,\stack))$ & $\hookrightarrow_{\id}$ & $(\gsto,\lsto,\stack,lb::b)$ &\\
$\mathcal{T}($\verb|block b|, $(\gsto,\lsto,\stack::v_0::v_i::v_j,lb::b::b_k))$ & $\hookrightarrow_{\id}$ & $(\gsto,\lsto,\stack::v_0::v_j)$ & $i=$node.nresults\\
$\mathcal{T}($\verb|loop b|, $(\gsto,\lsto,\stack))$ & $\hookrightarrow_{\id}$ & $(\gsto,\lsto,\stack,lb::b)$ &\\
$\mathcal{T}($\verb|endLoop b|, $(\gsto,\lsto,\stack::v_0::v_i::v_j,lb::b::b_k))$ & $\hookrightarrow_{\id}$ & $(\gsto,\lsto,\stack)$ & $i=$node.nresults\\
$\mathcal{T}($\verb|if|, $(\gsto,\lsto,\stack::v_0))$ & $\hookrightarrow_{\id}$ & $(\gsto,\lsto,\stack)$ &\\
$\mathcal{T}($\verb|br_if b|, $(\gsto,\lsto,\stack::v_0))$ & $\hookrightarrow_{\id}$ & $(\gsto,\lsto,\stack)$ &\\
$\mathcal{T}($\verb|br_table b|$^*$ \verb|b|$_N$, $(\gsto,\lsto,\stack::v_0))$ & $\hookrightarrow_{\id}$ & $(\gsto,\lsto,\stack)$ &\\
$\mathcal{T}($\verb|call x|, $(\gsto,\lsto,\stack::a_0::a_i))$ & $\hookrightarrow_{\id}$ & $(\gsto,\lsto,\stack::v_{j}^{?})$ & $j=$x.nresults\\
$\mathcal{T}($\verb|call_indirect t|, $(\gsto,\lsto,\stack::a_0::a_i::c))$ & $\hookrightarrow_{\id}$ & $(\gsto,\lsto,\stack::v_{j}^{?})$ & $j=$t.nresults\\
 \hline
\end{tabular}
\caption{Transfer Functions.}
\label{table:transfer_functions}
\end{table*}

\begin{algorithm}[t]
\small
\SetAlgoLined
\SetKwComment{Comment}{}{}
\KwIn{\textit{Entry Node}: Node Instructions of function $f$}
\KwResult{Map (instruction, rdef) for every instruction.}

 $DEP =$ \{ \\
\quad    $g$ \Comment*[l]{\hspace{10em}$\triangleright$ Map for global definitions.}
\quad    $l$ \Comment*[l]{\hspace{10.18em}$\triangleright$ Map for local definitions.}
\quad    $st := \emptyset$ \Comment*[l]{\hspace{7.5em}$\triangleright$ Lists of sets of definitions.}
\quad    $lb := \emptyset$ \Comment*[l]{\hspace{7.55em}$\triangleright$ Lists of labels.}
\} \;
 $ST = \{(\text{Insts},DEP)\}$  \Comment*[l]{\hspace{1.3em} $\triangleright$ Stack (node, rdef) (LIFO)}
 $res = \emptyset$  \Comment*[l]{\hspace{8.15em}$\triangleright$ Map (instruction, inDep) to return.}
\While(){$ST$ not Empty}{
    ($inst,inDep$) $:=$ Pop($ST$)\;
    $DEP := \mathcal{T}(inst,inDep)$\;
    \uIf{instType = Loop \textbf{and} ($DEP \cup res[inst]$) = $res[inst]$}{
        \textbf{continue}\;
    }\Else{
        $res[inst] := DEP$\; 
        \ForEach(){child $\in$ children($inst$, CFG)} {
            Push($ST$, (child, $RD$))\;
        }
    }
 }
 \Return $res$\;
 \caption{DDG data flow generation for function $f$.}
 \label{algorithm:pdg}
\end{algorithm}
\FloatBarrier

\clearpage
\subsection{Queries}
\subsubsection{Use-after-free}

\begin{figure} [h!]
	\centering
	\small
	\begin{cpp}[numbers=left, frame=l]
void VulnerabilityChecker::UseAfterFree() {
    for (auto func : Query::functions()) {
        for (auto const& item : config[CONTROL_FLOW]) {
            std::string source = item[SOURCE];
            std::string dest = item[DEST];
            auto sourcePredicate = Predicate().instType(InstType::Call).label(source);
            auto callSourceInsts = Query::instructions({func}, sourcePredicate);

            for (Node* callSource : callSourceInsts) {
                auto ddgEdgeCond = Query::ddgEdge(callSource->label(), DDGType::Function);

                Node* destNode = nullptr;
                auto destPredicate = Predicate()
                        .type(NodeType::Instruction)
                        .instType(InstType::Call)
                        .label(dest)
                        .EXEC(destNode = node)
                        .reaches(callSource, destNode, ddgEdgeCond);

                auto destInsts = Query::BFS({callSource}, destPredicate, Query::CFG_EDGES);
                for (Node* callDest : destInsts) {
                    Node* inst = nullptr;
                    auto uafPredicate = Predicate()
                            .inDDGEdge(callSource->label(), DDGType::Function)
                            .EXEC(inst = node)
                            .reaches(callSource, inst, ddgEdgeCond);
                    Node* parent = nullptr;
                    auto uafInst = NodeStream(callDest)
                            .BFS(uafPredicate, Query::CFG_EDGES, 1)
                            .filterOut(Predicate()
                                           .instType(InstType::Call)
                                           .label(dest)
                                           .Or()
                                           .EXEC(parent = node->getParent(0))
                                           .TEST(parent->instType() == InstType::Call && parent->label() == dest))
                            .findFirst();

                    if (uafInst.isPresent()) {
                        std::stringstream desc;
                        desc << "Value from call " << callSource->label()
                             << " used after call to " << dest;
                        vulns.emplace_back(VulnType::UaF, func->name(),
                                           uafInst.get()->label(), desc.str());
                    }
                }
            }
        }
    }
}


\end{cpp}
\caption{Use-after-free in C++.}
\label{fig:uaf_cpp}
\end{figure}

\begin{figure} [h!]
	\centering
	\small
	\begin{wql}[numbers=left, frame=l]
foreach func in functions():
    nodes := [n in instructions(func) : (n.instType = "Call") && (n.label in config["pairMalloc"])];
    foreach callMalloc in nodes:
        descendants := [n in descendantsCFG(callMalloc) : 
                        (n.instType = "Call") && (n.label = config["pairMalloc"][callMalloc.label]) && 
                        reachesDDG(callMalloc, n, "Function", callMalloc.label)];
        foreach callFree in descendants:
            uafs := [n in descendantsCFG(callFree) : reachesDDG(callMalloc, n, "Function", callMalloc.label)];
            if (!uafs.empty()):
                vulnerability("Use after free", func.name, callFree.label);

\end{wql}
\caption{Use-after-free in WQL.}
\label{fig:uaf_wql}
\end{figure}

\begin{figure} [h!]
	\centering
	\small
	\begin{neo4j}[numbers=left, frame=l]
MATCH (f:Function)-[:AST*1..]->(i:Instruction)
WHERE i.instType="Call"
	AND i.label IN mallocs
    
WITH * MATCH (i)-[:CFG*1..]->(free:Instruction)
WHERE free.instType="Call" AND free.label IN frees
    AND (i)-[:DDG*1.. {ddgType:"Function", label:i.label}]->(free)

WITH * MATCH (free)-[:CFG*1..]->(uaf:Instruction)
WHERE (i)-[:DDG*1.. {ddgType:"Function", label:i.label}]->(uaf)

RETURN uaf.label as caller, f.name as function;

\end{neo4j}
\caption{Use-after-free in Neo4j.}
\label{fig:uaf_wql}
\end{figure}

\begin{figure} [h!]
	\centering
	\small
	\begin{datalog}[numbers=left, frame=l]
uaf(FUNC_NAME, SOURCE, SINK, Z) :-
    uafSourceSink(SOURCE, SINK), call(X, SOURCE, _, _), reachesFunc(FUNC_NAME, X),
    call(Y, SINK, _, _), ddgEdge(_, Y, SOURCE, "Function", _), reachesDDG(X, Y, SOURCE, "Function", _),
    reaches(X, Y, "CFG"),
    instruction(Z, _), ddgEdge(_, Z, SOURCE, "Function", _), reachesDDG(X, Z, SOURCE, "Function", _),
    reaches(Y, Z, "CFG")
\end{datalog}
\caption{Use-after-free in Datalog.}
\label{fig:uaf_datalog}
\end{figure}

\FloatBarrier
\clearpage

\subsubsection{Tainted function to function}
\begin{figure} [h!]
	\centering
	\small
	\begin{cpp}[numbers=left, frame=l]
void VulnerabilityChecker::TaintedFuncToFunc() {
    std::set<std::string> sources = config.at(SOURCES);
    std::set<std::string> sinks = config.at(SINKS);

    for (auto func : Query::functions()) {
        auto query = Query::instructions({func}, [&](Node* node) {
            if (node->instType() == InstType::Call &&
                sinks.count(node->label()) == 1) {
                auto ddgEdges = node->inEdges(EdgeType::DDG);
                return Query::containsEdge(ddgEdges, [&](Edge* e) {
                    if (e->ddgType() == DDGType::Function &&
                        sources.count(e->label()) == 1) {
                        std::stringstream desc;
                        desc << "Source " << e->label() << " reaches sink "
                             << node->label();
                        vulns.emplace_back(VulnType::Tainted, func->name(),
                                           node->label(), desc.str());
                        return true;
                    }
                    return false;
                });
            }
            return false;
        });
    }
}

\end{cpp}
\caption{Tainted source-to-sink in C++.}
\label{fig:taintedff_cpp}
\end{figure}

\begin{figure} [h!]
	\centering
	\small
	\begin{wql}[numbers=left, frame=l]
foreach func in functions(): 
    sinkCalls := [n in instructions(func) : n.instType = "Call" && n.label in sinks && 
                                            !([e in n.inEdges : e.type = "DDG" && e.ddgType = "Function" && 
                                            e.label in sources].empty())]; 
    foreach sink in sinkCalls:
        vulnerability("Tainted", func.name, sink.label);
        
    
\end{wql}
\caption{Tainted source-to-sink in WQL.}
\label{fig:taintedff_cpp}
\end{figure}

\begin{figure} [h!]
	\centering
	\small
	\begin{neo4j}[numbers=left, frame=l]
MATCH (f:Function)-[:AST*1..]->(sink:Instruction)
WHERE sink.instType="Call" AND sink.label IN sinks

WITH * MATCH (src:Instruction)-[:DDG*1..]->(sink)
WHERE src.instType="Call" AND src.label IN sources
    AND (source_call)-[:DDG*1.. {ddgType:"Function", label:src.label}]->(sink)
RETURN src.label as source, sink.label as sink, 
    f.name as function;
\end{neo4j}
\caption{Tainted source-to-sink in Neo4j.}
\label{fig:taintedff_neo4j}
\end{figure}

\begin{figure} [h!]
	\centering
	\small
	\begin{datalog}[numbers=left, frame=l]
taintedFuncToFunc(FUNC_NAME, Y, SINK) :-
    sources(SOURCE), call(X, SOURCE, _, _), reachesFunc(FUNC_NAME, X),
    sinks(SINK), call(Y, SINK, _, _), reachesDDG(X, Y, "Function", SOURCE)
\end{datalog}
\caption{Tainted source-to-sink in Datalog.}
\label{fig:taintedff_neo4j}
\end{figure}

\FloatBarrier

\subsubsection{Buffer Overflow loops}
\begin{figure*}[h!]
    \centering
    \begin{wql}[numbers=left, frame=l]
foreach func in functions():
    loops := [n in instructions(func) : n.instType = "Loop"];
    foreach loop in loops:
        insts := descendantsAST(loop);
        descendants := nil;
        vars := List();
        stores := [n in insts : n.instType = "Store" && !(descendants := descendantsAST(n).empty() &&
                        ![child in descendants : child.instType = "Binary" && child.opcode = "i32.add"].empty() &&
                        ![child in descendants : (child.instType = "LocalGet" || child.instType = "LocalTee") && vars.append(child.label).empty()]];

        foreach var in vars:
            nodes := [n in insts : n.instType = "BrIf" && 
                                [descendant in descendantsAST(n) : descendant.instType = "Compare" && 
                                            [child in descendantsAST(descendant) : (child.instType = "LocalGet" && child.label = var) ||
                                                                                    (child.instType = "LocalTee" && child.label = var)]]];
            if (!nodes.empty():
                vulnerability("BO Loops", func.name, loop.label);

    \end{wql}
    
    \caption{Buffer overflow loops in WQL.}
    \label{fig:boLoops_wql}
\end{figure*}

\begin{figure*}[h!]
    \centering
    \begin{datalog}[numbers=left, frame=l]
bufferOverflow(FUNC_NAME, VAR, LOOP) :-
    loop(L, LOOP, _), reachesFunc(FUNC_NAME, L),
    binary(B, "i32.add"), reaches(L, B, "AST"),
    ddgEdge(X, B, VAR, "Local", _, _, _), ddgEdge(_, B, _, "Const", _, _, _),
    store(S, _), reaches(S, B, "AST"),
    brIf(BR, _), reaches(L, BR, "AST"),
    compare(COMP, _), reaches(BR, COMP, "AST"),
    ddgEdge(_, COMP, VAR, "Local", _, _, _).

    \end{datalog}
    
    \caption{Buffer overflow loops in Datalog.}
    \label{fig:boLoops_datalog}
\end{figure*}

\begin{figure*}[h!]
    \centering
    \begin{cpp}[numbers=left, frame=l]
for (auto func : Query::functions() {
    auto loops = Query::instructions({func}, Predicate().instType(InstType::Loop);

    for (Node* loop : loops) {
        auto insts = Query::BFS({loop}, Query::ALL_INSTS, Query::AST_EDGES);

        std::set<std::string> vars;
        auto stores = NodeStream(insts)
                .filter(Predicate().instType(InstType::Store)
                .child(0)
                .filter(Predicate()
                            .instType(InstType::Binary)
                            .opcode(Opcode::I32Add)
                .children(Query::AST_EDGES)
                .filter(Predicate()
                            .instType(InstType::LocalGet)
                            .Or()
                            .instType(InstType::LocalTee)
                .forEach([&](Node* node) { vars.insert(node->label(); })
                .toNodeSet();

        for (auto var : vars) {
            auto brIfsNotEq = NodeStream(insts)
                    .filter(Predicate().instType(InstType::BrIf)
                    .child(0)
                    .filter(Predicate().instType(InstType::Compare)
                    .children()
                    .filter(Predicate()
                                .instType(InstType::LocalGet)
                                .label(var)
                                .Or()
                                .instType(InstType::LocalTee)
                                .label(var)
                    .toNodeSet();

            if (brIfsNotEq.empty() {
                std::stringstream desc;
                desc << "In loop " << loop->label() << ":";
                desc << " a buffer is assigned without bound check.";
                vulns.emplace_back(VulnType::BufferOverflow, func->name(),
                                   "", desc.str();
                break;
            }
        }
    }
}

    \end{cpp}
    
    \caption{Buffer overflow loops in C++.}
    \label{fig:boLoops_cpp}
\end{figure*}

\begin{figure*}[h!]
    \centering
    \begin{neo4j}[numbers=left, frame=l]
// Get loops
MATCH (f:Function)-[:AST*1..]->(loop:Instruction)
WHERE loop.instType="Loop"

// Get stores and vars
WITH *
MATCH (loop)-[:AST*1..]->(store:Instruction)-[argEdge:AST]->(storeArg:Instruction)-[:AST]->(var:Instruction)
WHERE store.instType="Store"
    AND argEdge.arg=0
    AND storeArg.instType="Binary"
    AND storeArg.opcode="i32.add"
    AND (var.instType="LocalGet" OR var.instType="LocalTee")

// Get add instructions in function
WITH *
MATCH (f)-[:AST*1..]->(add:Instruction)
WHERE add.instType="Binary"
    AND add.opcode="i32.add"
    
// Check if var is being incremented with a constant
WITH *
MATCH (childConst:Instruction)<-[:AST]-(add)-[:AST]->(childLocal:Instruction)
WHERE childConst.instType="Const"
    AND childLocal.label=var.label
    AND (childLocal.instType="LocalGet" OR childLocal.instType="LocalTee")

// Check if breaks depend on store vars
WITH *
OPTIONAL MATCH path=(loop)-[:AST*1..]->(brIf:Instruction)-[:AST]->(compare:Instruction)-[:AST*1..]->(local:Instruction)
WHERE brIf.instType="BrIf"
    AND compare.instType="Compare"
    AND local.label=var.label
    AND (var.instType="LocalGet" OR var.instType="LocalTee")

WITH path, f, loop
RETURN DISTINCT
CASE path WHEN path=null THEN f.name END AS function,
CASE path WHEN path=null THEN loop.label END AS loop;

    \end{neo4j}
    
    \caption{Buffer overflow loops in Neo4j.}
    \label{fig:boLoops_neo4j}
\end{figure*}

\FloatBarrier

\subsection{Query Infrastructure}
\subsubsection{Native}
\begin{table}[h!]
    \centering
    \small
    \begin{tabular}{llc}\hline
        \textcolor{blue}{\texttt{NodeSet}} & \texttt{\textcolor{blue}{functions}()} & $O(1))$\\
         & Returns all the function nodes. &\\ \hline
         \textcolor{blue}{\texttt{NodeSet}} & \texttt{\textcolor{blue}{child}(\textcolor{blue}{NodeSet} set, \textcolor{blue}{Index} n, const \textcolor{blue}{EdgeType} type)} & $O(V))$\\
         & Returns the n'th child according to the edge type. &\\ \hline
         \texttt{\textcolor{blue}{NodeSet}} & \texttt{\textcolor{blue}{children}(\textcolor{blue}{NodeSet} set, \textcolor{blue}{EdgeCondition}\& cond)} & $O(V))$\\
         & Returns all the children following the edges according to an edge condition. &\\ \hline
         \texttt{\textcolor{blue}{NodeSet}} & \texttt{\textcolor{blue}{parents}(\textcolor{blue}{NodeSet} set, \textcolor{blue}{EdgeCondition}\& cond)} & $O(V))$\\
         & Returns all the parents following the edges according to an edge condition. &\\ \hline
         \texttt{\textcolor{blue}{NodeSet}} & \texttt{\textcolor{blue}{filter}(\textcolor{blue}{NodeSet} set, \textcolor{blue}{Predicate}\& pred)} & $O(V))$\\
         & Filters the node set according to the given predicate. &\\ \hline
          \texttt{\textcolor{blue}{EdgeSet}} & \texttt{\textcolor{blue}{filterEdges}(\textcolor{blue}{EdgeSet} set, const \textcolor{blue}{EdgeCondition}\& cond)} & $O(E))$\\
          & Filters the edge set according to the given predicate. &\\ \hline
         \texttt{\textcolor{blue}{NodeSet}} & \texttt{\textcolor{blue}{descendants}(\textcolor{blue}{NodeSet} set, \textcolor{blue}{Predicate}\& pred, \textcolor{blue}{EdgeCondition}\& cond)} & $O((N+E)/V))$\\
         & Returns all descendants nodes following an edge condition and a predicate.&\\ \hline
         \texttt{\textcolor{blue}{NodeSet}} & \texttt{\textcolor{blue}{instructions}(\textcolor{blue}{NodeSet} set, \textcolor{blue}{Predicate}\& pred)} & $O(N/V))$\\
         & Returns all instruction nodes from the function nodes in the set.&\\ 
     \hline
    \end{tabular}
    \caption{Basic query API.}
    \label{table:query_api}
\end{table}
\begin{table}[h!]
    \centering
    \small
    \begin{tabular}{llc}\hline
        \texttt{\textcolor{blue}{Predicate}\&} & \texttt{\textcolor{blue}{$<$keyProperty$>$}(\textcolor{blue}{Value} value, \textcolor{blue}{bool} equal = true)} & $O(1))$\\
         & Compares the value from property $<$keyProperty$>$ against the provided value. &\\ \hline
        \texttt{\textcolor{blue}{Predicate}\&} & \texttt{\textcolor{blue}{inEdge}(\textcolor{blue}{EdgeType} type, \textcolor{blue}{std::string} label, 
         \textcolor{blue}{bool} equal = true)} & $O(E))$\\
         & Check if node contains and incoming edge with given type and label. &\\ \hline
         \texttt{\textcolor{blue}{Predicate}\&} &  \texttt{\textcolor{blue}{inDDGEdge}(\textcolor{blue}{DDGType} type, \textcolor{blue}{std::string} label, 
         \textcolor{blue}{bool} equal = true)} & $O(E))$\\
         & Check if node contains and incoming DDG edge with given DDG type and label. &\\ \hline
         \texttt{\textcolor{blue}{Predicate}\&} &  \texttt{\textcolor{blue}{outEdge}(\textcolor{blue}{EdgeType} type, \textcolor{blue}{std::string} label, 
         \textcolor{blue}{bool} equal = true)} & $O(E))$\\
         & Check if node contains and out-coming edge with given type and label. &\\ \hline
         \texttt{\textcolor{blue}{Predicate}\&} &  \texttt{\textcolor{blue}{outDDGEdge}(\textcolor{blue}{DDGType} type, \textcolor{blue}{std::string} label, 
         \textcolor{blue}{bool} equal = true)} & $O(E))$\\
         & Check if node contains and ot-coming DDG edge with given DDG type and label. &\\ \hline
         \texttt{\textcolor{blue}{Predicate}\&} &  \texttt{\textcolor{blue}{reachesIn}(\textcolor{blue}{Node*} src, \textcolor{blue}{EdgeCondition}\& cond)} & $O(N+E))$\\
         & Check if there is a patch from src following an edge condition. &\\ \hline
         \texttt{\textcolor{blue}{Predicate}\&} &  \texttt{\textcolor{blue}{reachesOut}(\textcolor{blue}{Node*} dest, \textcolor{blue}{EdgeCondition}\& cond)} & $O(N+E))$\\
         & Check if there is a patch to dest following an edge condition. &\\ \hline
         \texttt{\textcolor{blue}{Predicate}\&} &  \texttt{\textcolor{blue}{test}(\textcolor{blue}{std::function$<$bool(Node*))$>$} func)} & $O(\verb|func|))$\\
         & Tests the condition given by lambda function func. &\\ \hline
    \end{tabular}
    \caption{Predicate API.}
    \label{table:predicate_api}
\end{table}

\subsection{Evaluation Tables}

\begin{table*}[t]
  \centering
  {\footnotesize
  \resizebox{\textwidth}{!}{
    \begin{tabular}{lrrrrrrrr}
    \toprule
    \textbf{Binary} & \multicolumn{1}{l}{\textbf{Source}} & \multicolumn{1}{l}{\textbf{Instruct. (k)}} & \multicolumn{1}{l}{\textbf{Size (KiB)}} & \multicolumn{1}{l}{\textbf{Nodes}} & \multicolumn{1}{l}{\textbf{Edges}} & \multicolumn{1}{l}{\textbf{Memory (MiB)}} & \multicolumn{1}{c}{\textbf{Time}} & \multicolumn{1}{l}{\textbf{Exported (MiB)}} \\
    \midrule
    \rowcolor[rgb]{ .929,  .929,  .929} 500.perlbench\_r & \multicolumn{1}{l}{C} & 837.8k & 1,964 & 879.0k & 2.5M  & 175.87 & 45.34s & 14.00 \\
    502.gcc\_r & \multicolumn{1}{l}{C} & 2.9M  & 6,964 & 3.1M  & 9.6M  & 642.24 & 4min 15.97s & 53.00 \\
    \rowcolor[rgb]{ .929,  .929,  .929} 505.mcf\_r & \multicolumn{1}{l}{C} & 27.4k & 56    & 30.0k & 89.0k & 6.14  & 1.20s & 0.48 \\
    508.namd\_r & \multicolumn{1}{l}{C++} & 323.0k & 636   & 343.0k & 813.0k & 64.05 & 7.62s & 4.70 \\
    \rowcolor[rgb]{ .929,  .929,  .929} 510.parest\_r & \multicolumn{1}{l}{C++} & 1.0M  & 2,190 & 1.1M  & 3.5M  & 226.47 & 33.51s & 19.00 \\
    511.povray\_r & \multicolumn{1}{l}{C++} & 385.4k & 909   & 406.5k & 1.4M  & 90.06 & 4min 19.76s & 7.30 \\
    \rowcolor[rgb]{ .929,  .929,  .929} 519.lbm\_r & \multicolumn{1}{l}{C} & 13.4k & 29    & 14.6k & 55.2k & 3.36  & 1.39s & 0.26 \\
    520.omnetpp\_r & \multicolumn{1}{l}{C++} & 619.3k & 1,524 & 658.0k & 4.3M  & 205.01 & 23.67s & 19.00 \\
    \rowcolor[rgb]{ .929,  .929,  .929} 523.xalancbmk\_r & \multicolumn{1}{l}{C++} & 1.5M  & 3,404 & 1.5M  & 13.7M & 587.35 & 58.53s & 58.00 \\
    525.ldecod\_r & \multicolumn{1}{l}{C} & 233.0k & 476   & 224.0k & 624.2k & 44.67 & 6.44s & 3.40 \\
    \rowcolor[rgb]{ .929,  .929,  .929} 525.x264\_r & \multicolumn{1}{l}{C} & 283.6k & 592   & 282.1k & 864.1k & 58.65 & 8.88s & 4.60 \\
    526.blender\_r & \multicolumn{1}{l}{C++} & 3.2M  & 9,944 & 3.4M  & 44.1M & 1,735.99 & 3min 45.86s & 184.00 \\
    \rowcolor[rgb]{ .929,  .929,  .929} 531.deepsjeng\_r & \multicolumn{1}{l}{C} & 53.0k & 112   & 56.6k & 158.0k & 11.30 & 2.15s & 0.88 \\
    538.imagick\_r & \multicolumn{1}{l}{C} & 517.5k & 1,216 & 552.5k & 1.6M  & 112.10 & 35.74s & 8.90 \\
    \rowcolor[rgb]{ .929,  .929,  .929} 541.leela\_r & \multicolumn{1}{l}{C++} & 118.8k & 272   & 135.5k & 384.0k & 27.23 & 3.08s & 2.10 \\
    544.nab\_r & \multicolumn{1}{l}{C} & 55.6k & 122   & 60k   & 172.6k & 12.13 & 2.85s & 0.94 \\
    \rowcolor[rgb]{ .929,  .929,  .929} 557.xz\_r & \multicolumn{1}{l}{C} & 53.3k & 136   & 57.8k & 185.8k & 12.29 & 10.37s & 0.94 \\
    \midrule
    \multicolumn{2}{l}{\textbf{Average} \textit{per binary}}       & \textbf{713.0k} & \textbf{1,797} & \textbf{747.6k} & \textbf{4.9M} & \textbf{236.17} & \textbf{57.79s} & \textbf{22.44} \\
    \textbf{Total} &       & \textbf{12.1M} & \textbf{30,547} & \textbf{12.7M} & \textbf{83.9M} & \textbf{4014.91} & \textbf{16min 22.36s} & \textbf{381.50} \\
    \bottomrule
    \end{tabular}}
    }
  \caption{CPG generation times and information from SPEC CPU 2017 binaries.}
  \label{tab:spec2017}
  \vspace{-0.3cm}
\end{table*}

\begin{table*}[t]
  \centering
  {\footnotesize
  \resizebox{\textwidth}{!}{
    \begin{tabular}{l|rrrrrrrrrr|r}
    \hline
    \textbf{Binary} & \textbf{1} & \textbf{2} & \textbf{3} & \textbf{4} & \textbf{5} & \textbf{6} & \textbf{7} & \textbf{8} & \textbf{9} & \textbf{10} & \textbf{Total} \\
    \hline
500.perlbench\_r & \cellcolor{blue!15}5.72 & \cellcolor{blue!15}6.13 & \cellcolor{blue!15}5.53 & \cellcolor{blue!15}5.22 & \cellcolor{blue!15}4.41 & \cellcolor{blue!15}6.04 & 0.09 & \cellcolor{blue!30}27.18 & \cellcolor{blue!15}6.18 & \cellcolor{blue!15}5.66 & 72.17   \\
 502.gcc\_r & \cellcolor{blue!30}17.93 & \cellcolor{blue!30}22.93 & \cellcolor{blue!30}17.32 & \cellcolor{blue!30}16.59 & \cellcolor{blue!30}21.11 & \cellcolor{blue!30}17.70 & \cellcolor{blue!5}0.40 & \cellcolor{blue!30}89.02 & \cellcolor{blue!30}19.70 & \cellcolor{blue!30}27.35 & 250.05  \\
 505.mcf\_r & \cellcolor{blue!5}0.15 & \cellcolor{blue!5}0.17 & \cellcolor{blue!5}0.17 & \cellcolor{blue!5}0.16 & \cellcolor{blue!5}0.17 & \cellcolor{blue!5}0.22 & 0.02 & \cellcolor{blue!5}0.88 & \cellcolor{blue!5}0.15 & \cellcolor{blue!5}0.33 & 2.41    \\
 508.namd\_r     & \cellcolor{blue!15}1.96 & \cellcolor{blue!15}1.79 & \cellcolor{blue!15}2.21 & \cellcolor{blue!15}2.32 & \cellcolor{blue!15}2.16 & \cellcolor{blue!15}1.74 & 0.02 & \cellcolor{blue!15}7.36 & \cellcolor{blue!15}1.84 & \cellcolor{blue!15}3.25 & 24.64   \\
 510.parest\_r   & \cellcolor{blue!15}6.49 & \cellcolor{blue!15}6.76 & \cellcolor{blue!15}7.07 & \cellcolor{blue!15}7.29 & \cellcolor{blue!15}6.31 & \cellcolor{blue!15}6.95 & \cellcolor{blue!5}0.15 & \cellcolor{blue!30}37.23 & \cellcolor{blue!15}7.54 & \cellcolor{blue!30}10.73 & 96.52   \\
 511.povray\_r   & \cellcolor{blue!15}2.17 & \cellcolor{blue!15}2.29 & \cellcolor{blue!15}2.40 & \cellcolor{blue!15}2.80 & \cellcolor{blue!15}2.39 & \cellcolor{blue!15}2.77 & 0.06 & \cellcolor{blue!15}9.69 & \cellcolor{blue!15}2.27 & \cellcolor{blue!15}3.78 & 30.63   \\
 519.lbm\_r & 0.08 & 0.08 & 0.09 & 0.08 & 0.07 & \cellcolor{blue!5}0.10 & 0.02 & \cellcolor{blue!5}0.50 & 0.07 & \cellcolor{blue!5}0.23 & 1.33    \\
 520.omnetpp\_r  & \cellcolor{blue!15}4.60 & \cellcolor{blue!15}3.89 & \cellcolor{blue!15}3.71 & \cellcolor{blue!15}5.16 & \cellcolor{blue!15}5.82 & \cellcolor{blue!15}4.01 & \cellcolor{blue!5}0.18 & \cellcolor{blue!30}20.78 & \cellcolor{blue!15}4.06 & \cellcolor{blue!15}7.26 & 59.47   \\
 523.xalancbmk\_r & \cellcolor{blue!30}11.61 & \cellcolor{blue!30}12.90 & \cellcolor{blue!30}10.78 & \cellcolor{blue!30}13.14 & \cellcolor{blue!30}15.41 & \cellcolor{blue!30}13.15 & \cellcolor{blue!5}0.33 & \cellcolor{blue!30}65.28 & \cellcolor{blue!30}11.98 & \cellcolor{blue!30}16.58 & 171.17  \\
 525.ldecod\_r   & \cellcolor{blue!15}1.26 & \cellcolor{blue!15}1.10 & \cellcolor{blue!15}1.39 & \cellcolor{blue!15}1.34 & \cellcolor{blue!15}1.38 & \cellcolor{blue!15}1.43 & 0.03 & \cellcolor{blue!15}6.34 & \cellcolor{blue!15}1.32 & \cellcolor{blue!15}1.98 & 17.59   \\
 525.x264\_r     & \cellcolor{blue!15}1.64 & \cellcolor{blue!15}1.93 & \cellcolor{blue!15}1.66 & \cellcolor{blue!15}1.91 & \cellcolor{blue!15}1.46 & \cellcolor{blue!15}1.59 & 0.04 & \cellcolor{blue!15}8.65 & \cellcolor{blue!15}1.71 & \cellcolor{blue!15}2.29 & 22.87   \\
 526.blender\_r  & \cellcolor{blue!30}32.81 & \cellcolor{blue!30}41.00 & \cellcolor{blue!30}36.17 & \cellcolor{blue!30}32.18 & \cellcolor{blue!30}35.70 & \cellcolor{blue!30}41.28 & \cellcolor{blue!15}1.08 & \cellcolor{blue!50}194.12 & \cellcolor{blue!30}32.11 & \cellcolor{blue!30}41.53 & 487.98  \\
 531.deepsjeng\_r & \cellcolor{blue!5}0.30 & \cellcolor{blue!5}0.39 & \cellcolor{blue!5}0.39 & \cellcolor{blue!5}0.33 & \cellcolor{blue!5}0.28 & \cellcolor{blue!5}0.37 & 0.02 & \cellcolor{blue!15}1.89 & \cellcolor{blue!5}0.34 & \cellcolor{blue!5}0.88 & 5.19    \\
 538.imagick\_r  & \cellcolor{blue!15}2.81 & \cellcolor{blue!15}3.00 & \cellcolor{blue!15}2.96 & \cellcolor{blue!15}3.88 & \cellcolor{blue!15}3.98 & \cellcolor{blue!15}2.67 & 0.05 & \cellcolor{blue!30}14.72 & \cellcolor{blue!15}3.34 & \cellcolor{blue!15}4.70 & 42.11   \\
 541.leela\_r    & \cellcolor{blue!5}0.63 & \cellcolor{blue!5}0.71 & \cellcolor{blue!5}0.86 & \cellcolor{blue!5}0.59 & \cellcolor{blue!5}0.80 & \cellcolor{blue!5}0.73 & 0.06 & \cellcolor{blue!15}3.75 & \cellcolor{blue!5}0.82 & \cellcolor{blue!15}1.63 & 10.57   \\
 544.nab\_r & \cellcolor{blue!5}0.28 & \cellcolor{blue!5}0.35 & \cellcolor{blue!5}0.38 & \cellcolor{blue!5}0.41 & \cellcolor{blue!5}0.34 & \cellcolor{blue!5}0.29 & 0.02 & \cellcolor{blue!15}1.79 & \cellcolor{blue!5}0.31 & \cellcolor{blue!5}0.98 & 5.16    \\
 557.xz\_r & \cellcolor{blue!5}0.33 & \cellcolor{blue!5}0.30 & \cellcolor{blue!5}0.29 & \cellcolor{blue!5}0.39 & \cellcolor{blue!5}0.34 & \cellcolor{blue!5}0.31 & 0.02 & \cellcolor{blue!15}1.57 & \cellcolor{blue!5}0.40 & \cellcolor{blue!5}0.72 & 4.67    \\
    \hline
    \textbf{Average} \textit{per binary} & \textbf{\cellcolor{blue!15}5.34} & \textbf{\cellcolor{blue!15}6.22} & \textbf{\cellcolor{blue!15}5.49} & \textbf{\cellcolor{blue!15}5.52} & \textbf{\cellcolor{blue!15}6.01} & \textbf{\cellcolor{blue!15}5.96} & \textbf{\cellcolor{blue!5}0.15} & \textbf{\cellcolor{blue!30}28.87} & \textbf{\cellcolor{blue!15}5.54} & \textbf{\cellcolor{blue!15}7.64} & \textbf{76.74} \\
    \textbf{Total}  & \textbf{90.75} & \textbf{105.72} & \textbf{93.38} & \textbf{93.79} & \textbf{102.12} & \textbf{101.37} & \textbf{2.60} & \textbf{490.76} & \textbf{94.15} & \textbf{129.88} & \textbf{1304.52} \\
    \hline
    \end{tabular}}
    }
  \caption{\small WQL execution time for each query over the SPEC binaries (in seconds). Darker cells represent higher execution times.}
  \label{tab:queryExecTime}
  \vspace{-0.3cm}
\end{table*}


\end{document}